\documentclass[journal,twocolumn,final]{IEEEtran}

\usepackage{cite}
\usepackage{pbox}
\usepackage{pstricks}
\usepackage{graphicx}
\usepackage{amsmath}
\usepackage{amssymb}
\usepackage{amsthm}
\usepackage{mathrsfs}
\usepackage{mathtools}
\usepackage{subcaption}
\usepackage{tikz,pgfplots}
\pgfplotsset{compat=newest}
\usetikzlibrary{graphs,matrix,positioning,calc,fit,backgrounds,chains}
\usepackage[breakable]{tcolorbox}
\tcbuselibrary{breakable}
\usepackage{tikzscale}
\usepackage{ulem}
\let\oldbrace\{
\def\{{\oldbrace\kern0.5pt}

\def\Var{\mathop{\rm Var}\nolimits}%
\newcommand{\nn}{\nonumber}

\newcommand{\Cc}{\mathcal{C}}

\newcommand{\Nc}{\mathcal{N}}

\newcommand{\Pc}{\mathcal{P}}

\newcommand{\Sc}{\mathcal{S}}
\newcommand{\Tc}{\mathcal{T}}

\newcommand{\Xc}{\mathcal{X}}

\newcommand{\Ds}{\mathsf{D}}

\newcommand{\Xv}{\boldsymbol{X}}
\newcommand{\Yv}{\boldsymbol{Y}}
\newcommand{\Zv}{\boldsymbol{Z}}

\newcommand{\xv}{\boldsymbol{x}}
\newcommand{\yv}{\boldsymbol{y}}

\newcommand{\pen}{{P_e^{(n)}}}

\newcommand{\Yt}{{\tilde{Y}}}
\newcommand{\Zt}{{\tilde{Z}}}

\newcommand{\yt}{{\tilde{y}}}

\def\e{\epsilon}

\DeclareMathOperator\E{\sf E}
\let\P\relax
\DeclareMathOperator\P{\sf P}

\newtheorem{theorem}{Theorem}
\newtheorem{lemma}{Lemma}
\newtheorem{corollary}{Corollary}

\theoremstyle{definition}

\newtheorem{remark}{Remark}
\newtheorem{definition}{Definition}

\DeclareMathOperator*{\esssup}{\textrm{ess}\sup}
\DeclareMathOperator*{\esssupe}{\textrm{\em ess}\sup}

\newcommand{\FAR}{{\textsf{FAR}}}
\newcommand{\WADD}{\textsf{WADD}}
\newcommand{\SPRT}{\textsf{SPRT}}
\newcommand{\CS}{\textsf{CS}}
\newcommand{\SCS}{\textsf{SCS}}
\newcommand{\inn}{\textsf{in}}
\newcommand{\Ccn}{\Cc^{(n)}}

\IEEEoverridecommandlockouts

\begin{document}
\title{On the Fundamental Tradeoff of Joint Communication and Quickest Change Detection with State-Independent Data Channels}

\author{Daewon Seo,~\IEEEmembership{Member, IEEE,} and Sung Hoon Lim,~\IEEEmembership{Senior Member, IEEE}\\
	\thanks{ D.~Seo is with the Department of Electrical Engineering and Computer Science, Daegu Gyeongbuk Institute of Science and Technology (DGIST), Daegu 42988, South Korea (e-mail: dwseo@dgist.ac.kr). Sung Hoon Lim is with the School of Information Sciences, Hallym University, Chuncheon 24252, South Korea (e-mail: shlim@hallym.ac.kr).}}
	
\maketitle
\allowdisplaybreaks

\begin{abstract}
In this work, we take the initiative in studying the information-theoretic tradeoff between communication and quickest change detection (QCD) under an integrated sensing and communication setting. We formally establish a joint communication and sensing problem for the quickest change detection.
We assume a broadcast channel with a transmitter, a communication receiver, and a QCD detector in which only the detection channel is state dependent. For the problem setting, by utilizing constant subblock-composition codes and a modified CuSum detection rule, which we call subblock CuSum (SCS), we provide an inner bound on the information-theoretic tradeoff between communication rate and change point detection delay in the asymptotic regime of vanishing false alarm rate. We further provide a partial converse that matches our inner bound for a certain class of codes. This implies that the SCS detection strategy is asymptotically optimal for our codes as the false alarm rate constraint vanishes. We also present some canonical examples of the tradeoff region for a binary channel, a scalar Gaussian channel, and a MIMO Gaussian channel.
\end{abstract}
\begin{IEEEkeywords}
Integrated sensing and communication, information theory, quickest change detection, CuSum test, constant subblock-composition codes
\end{IEEEkeywords}

\section{Introduction} \label{sec:intro}
The capability to simultaneously perform communication and sensing tasks efficiently within limited resources is one of the key features envisioned for next-generation wireless networks~\cite{Bourdoux2020_2,Liu--Huang--Li--wan--Li--Han--Liu--Du--Tan--Lu--Shen--Colone--Chetty2022_2}. In contrast to conventional systems typically designed with separately dedicated resources and hardware, integrated sensing and communication (ISAC) systems aim to cohesively design both sensing and communication functionalities. This design approach enables the system to share common resources, such as frequency bands and hardware, in order to enhance efficiency and reduce costs~\cite{Liu--Masouros--Petropulu--Griffiths--Hanzo2020, Sturm--Wiesbeck2011}.  Henceforth, it is expected that ISAC will become a key technology in future wireless systems, supporting many important application scenarios such as device localization, autonomous vehicles, surveillance, and healthcare~\cite{Liu--Masouros--Petropulu--Griffiths--Hanzo2020, Akan--Arik2020, Liu--Cui--Masouros--Xu--Han--Eldar--Buzzi2022}.

In ISAC, the primary focus of research is on simultaneous communication of information and target sensing. Consequently, it is natural to explore the tradeoff between two key metrics; one related to data transmission and the other to sensing accuracy. For instance, the tradeoff between information rate for transmitting data reliably and mean-squared error (MSE) for estimating an unknown real-valued channel state is characterized in \cite{Sutivong--Chiang--Cover--Kim2005, Xiong--Liu--Cui--Yuan--Han--Caire2023}. In another instance, the tradeoff between information rate and error probability of detecting a discrete target, e.g., obstacle detection, is investigated in \cite{Joudeh--Willems2022, Ahmadipour--Kobayashi--Wigger--Caire2022, Chang--Wang--Erdogan--Bloch--2023, Wu--Joudeh2022}. In addition to these intrinsic ISAC metrics, the metric of secrecy is also considered in \cite{Gunlu--Bloch--Schaefer--Yener2023, Ren--Qiu--Xu--Ng2023}. Some surveys and overviews on the topics of ISAC are given in~\cite{Liu--Cui--Masouros--Xu--Han--Eldar--Buzzi2022, Liu--Huang--Li--wan--Li--Han--Liu--Du--Tan--Lu--Shen--Colone--Chetty2022, Cui--Liu--Jing--Mu2021, Zheng--Lops--Eldar--Wang2019}.

Building upon such previous studies, we introduce a new metric of ISAC that requires detecting changes or anomalies in the channel {\em as soon as possible} subject to some false alarm constraint. We note that the sole problem of detecting the change point in distribution from a sequence of observations (i.e., without a communication requirement) is alone an active field of research in the context of quickest change detection (QCD)~\cite{Lorden1971, Pollak1985, Lai1998, Veeravalli--Banerjee2014, Xie--Zou--Xie--Veeravalli2021}. In this work, our goal is to perform communication and QCD with a common signal and provide fundamental tradeoffs and design perspectives for joint communication and QCD.

One possible motivating scenario for this requirement could arise from a situation in which a device equipped with wireless communication capabilities, such as an unmanned aerial vehicle (UAV), an autonomous vehicle, or a robot arm, must identify sudden obstacles in its trajectory. In another scenario, base stations scattered in a wireless network can be utilized as a surveillance sensor network to detect abnormalities with minimum delay. In these instances, the transmitted signal for communication simultaneously serves as an active sensing signal for detecting anomalies. Thus, the policy for designing a transmission signal must be jointly optimized to provide the best tradeoff between communication rate and detection delay. 

However, there are some inherent fundamental challenges to achieving such goals. On the one hand, results on capacity-achieving codes for communication over a noisy channel suggest that the signals should be designed in units of long blocks that are distinguishable from other competing codewords when decoded over observation sequences. On the other hand, the codewords should also act as potent active sensing signals that produce the most distinctive observations that allow detection of the change point with minimum delay.
\begin{figure}[!t]
\begin{center}
\resizebox{!}{10.2em}{
\begin{tikzpicture}[font=\large,node distance=.6cm and 1cm, start chain]
     \tikzstyle{rect}=[draw=black, 
                   rectangle, 
                   text opacity=1,
                   minimum width=50pt, 
                   minimum height = 25pt, 
                   align=center]
  \node[rect] (encoder) {Encoder};
  \node[rect,above right=0.4cm and 1.6cm of encoder] (com_chn) {$p_{\tilde{Y}|X}$};
  \node[rect,below right=0.4cm and 1.6cm of encoder] (qcd_chn) {$p_{Y|X, S}$};
  \node[rect, right=of com_chn] (decoder) {Decoder};
  \node[rect, right=of qcd_chn] (detector) {Detector};
\draw[-stealth] (encoder.east) -- node[above] {$X^n$} +(1,0) |- (com_chn.west);
\draw[-stealth] (encoder.east) -- +(1,0) |- (qcd_chn.west);
\draw[-stealth] (com_chn.east) -- node[above] {$\tilde{Y}^n$}(decoder.west);
\draw[-stealth] (qcd_chn.east) -- node[above] {$Y^n$}(detector.west);
\draw[stealth-] (encoder.west) --+ (-5mm,0) node[left] {$M$};
\draw[-stealth] (decoder.east) --+ (5mm,0) node[right] {$\hat{M}$};
\draw[-stealth] (detector.east) --+ (5mm,0) node[right] {$N$};
\draw[stealth-] (detector.south) --+ (0,-5mm) node[below] {$M$};
\end{tikzpicture}}
\end{center}
\caption{Problem model of joint communication and QCD.}\label{fig:model1}
\end{figure}
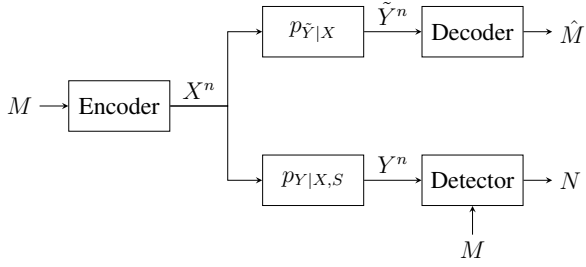
To put the problem into perspective, we consider a broadcast channel as shown in Fig.~\ref{fig:model1} that consists of one encoder and two receivers. The encoder transmits an $n$ length codeword $x^n$ through a state-dependent memoryless broadcast channel $p(y^n, \yt^n|x^n, s^n) = \prod_{i=1}^n p(y_i|x_i,s_i) p(\yt_i|x_i)$, where $s^n$ is a state sequence that starts from an initial state but at some unknown change point, transitions to another state. Further, $\yt^n, y^n$ are the sequences of observations for the information decoder and QCD detector, respectively. The decoder recovers a message sent from the encoder, $\hat{M}$, and the QCD detector detects the change point with minimum delay by establishing a stopping time, $N$, in an online manner. 
The model we consider is the bistatic ISAC model in the literature~\cite{Xiong--Liu--Cui--Yuan--Han--Caire2023, Chang--Wang--Erdogan--Bloch--2023} with the additional assumption that the communication channel is independent of the state. The considered model has two notable limitations, firstly it does not include feedback of the detected state (cf. monostatic model), and the communication channel is independent of the state. However, it includes the important feature that a single transmitted signal must serve both as an
active sensing signal and as a data-bearing communication signal. Thus, our work focuses on providing information-theoretic guidelines on how to design codewords that are simultaneously useful for both communication and QCD.

We also assume that the message is known at the QCD detector as side information. This setup can serve as a canonical model, such as in a scenario where two base stations with a wireline link perform QCD in a wireless network, in which the message is sent through the wireline link as side information, e.g., \cite{Joudeh--Willems2022, Xiong--Liu--Cui--Yuan--Han--Caire2023}. This can be also thought of as a monostatic model where the encoder and the QCD detector are equipped on a single device, but feedback codes are not used. For this case, $p_{Y|X,S}$ is the deflection channel from a target.

\begin{figure}[!t]
\begin{center}
\resizebox{!}{9em}{
\begin{tikzpicture}[font=\large, node distance=.6cm and 0.5cm, start chain]
     \tikzstyle{rect}=[draw=black, 
                   rectangle, 
                   text opacity=1,
                   minimum width=20pt, 
                   minimum height = 20pt, 
                   align=center]
  \node (x1) {$x_1$};
  \node[right=of x1] (dots1) {$\cdots$};

  \node[right=of dots1] (xvm1) {$x_{\nu-1}$};
  \node[right=of xvm1] (xv) {$x_{\nu}$};
  \node[right=of xv] (xvp1) {$x_{\nu+1}$};
  \node[right=of xvp1] (dots2) {$\cdots$};
  \node[right=of dots2] (xn) {$x_{n}$};
  
  \node[rect, below=of x1] (chn1) {$p^{(0)}$};
  \node[below=of dots1] (dots21) {$\cdots$};
  \node[rect, below=of xvm1] (chn3) {$p^{(0)}$};
  \node[rect, below=of xv] (chn4) {$p^{(1)}$};
  \node[rect, below=of xvp1] (chn5) {$p^{(1)}$};
  \node[below=of dots2] (dots22) {$\cdots$};
  \node[rect, below=of xn] (chn7) {$p^{(1)}$};

  \node[below=of chn1] (y1) {$y_1$};
  \node[below=of dots21] (dots31) {$\cdots$};
  \node[below=of chn3] (yvm1) {$y_{\nu-1}$};
  \node[below=of chn4] (yv) {$y_\nu$};
  \node[below=of chn5] (yvp1) {$y_{\nu+1}$};
  \node[below=of dots22] (dots32) {$\cdots$};
  \node[below=of chn7] (yn) {$y_n$};

 \draw[-stealth] (x1.south) -- (chn1.north);
 \draw[-stealth] (xvm1.south) -- (chn3.north);
 \draw[-stealth] (xv.south) -- (chn4.north);
 \draw[-stealth] (xvp1.south) -- (chn5.north);
 \draw[-stealth] (xn.south) -- (chn7.north);

 \draw[-stealth] (chn1.south) -- (y1.north);
 \draw[-stealth] (chn3.south) -- (yvm1.north);
 \draw[-stealth] (chn4.south) -- (yv.north);
 \draw[-stealth] (chn5.south) -- (yvp1.north);
 \draw[-stealth] (chn7.south) -- (yn.north);
 \draw[dashed] ([xshift=0.2cm, yshift=0.5cm] xvm1.east) -- ([xshift=0.2cm, yshift=-1cm] yvm1.east);

 \node[below left= 0.5cm and -0.7cm of yvm1, anchor=east] (dist1) {$p^{(0)}(y|x)$};
 \node[below right= 0.5cm and -0.5cm of yv, anchor=west] (dist2) {$p^{(1)}(y|x)$};

 \end{tikzpicture}}
\end{center}
\caption{The distribution of the QCD channel changes at some turning point $\nu$ from the baseline distribution $p^{(0)}$ to the abnormal distribution $p^{(1)}$.}\label{fig:qcd}
\end{figure}
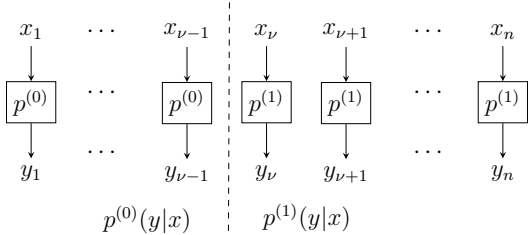

The state sequence represents when the change occurs in the distribution of the QCD channel. For example, consider a binary state channel shown in Fig.~\ref{fig:qcd} where $p^{(0)}(y|x) := p(y|x,s=0)$ and $p^{(1)}(y|x) := p(y|x,s=1)$. We assume that the QCD channel stays in its base state $s=0$ until (just before) a change point $\nu$, and from point $\nu$ and after, the channel stays in state $s=1$. The goal of the QCD detector is to provide a change point estimate $N$ with a minimum delay compared to the true change point $\nu$ under a false alarm constraint. 

This joint communication and QCD setting has several distinct features that should be noted. In the context of QCD, this problem is related to the active sensing QCD problem studied in~\cite{Gopalan--Lakshminarayanan--Saligrama2021, Veeravalli--Fellouris--Moustakides2024}, where the main objective is to design a control signal so that observations include as much information as possible for quickly detecting the change. The major distinction in our case is that the control signal must be simultaneously useful for communication. In the other perspective, relating to the field of communications and information theory, our work is also related to state amplification and state estimation in a state-dependent communication channel~\cite{Kim--Sutivon--Cover2008,Choudhuri--Kim--Mitra2020,Zhang--Vedantam--Mitra2011}. The major distinction in our case is that, first of all, we specialize our attention to a particular distortion criteria, namely, change point detection {\em delay}. Furthermore, we follow the minimax approach commonly used in the QCD literature and analyze the worst-case delay over all codewords. This is in contrast to many ISAC scenarios which consider average distortion measures over all codewords, e.g., \cite{Ahmadipour--Kobayashi--Wigger--Caire2022}. As the name suggests, QCD applications are more likely to exhibit some degree of urgency in anomaly detection, which makes worst-case delay the proper criteria. However, to match the worst-case delay criteria, {\em all} codewords must be good for sensing since the codewords are randomly chosen by the message. This disqualifies most standard random i.i.d.~generated codebooks since even with vanishing probability, the existence of ``bad'' codewords for sensing will completely dominate the worst-case performance metric. Finally, another distinction comes from the criteria of minimizing delay, which suggests that one should design an online detector since performing an estimation after receiving the whole block would simply give the worst change point estimation delay.

In this work, we contribute to the problem of joint communication and QCD as follows. Firstly, we take the initializing steps towards addressing the problem and provide an information-theoretic formulation that integrates two heterogeneous criteria which can be readily extended to various scenarios. Most importantly, we provide an inner bound on the fundamental tradeoff between communication and QCD. The precise result is summarized in Thm.~\ref{thm:theorem1}, where the achievable communication rate and the asymptotic delay slope (with respect to the false alarm rate) are expressed in the form of mutual information and Kullback--Leibler (KL) divergence, respectively. The achievable region can be efficiently computed using the Blahut--Arimoto algorithm~\cite{Blahut1972, Arimoto1972}. For the achievability proof, we utilize constant subblock-composition codes (CSCCs) originally developed in~\cite{Tandon--Motani--Varshney2016}, where the composition of transmitted symbols in every subblock is identical, and our proposed subblock CuSum (SCS) test, a modified version of the standard CuSum test. As a result of CSCCs, almost every subblock of observations can be made independent and identically distributed by proper processing, which enables us to extend standard QCD analysis techniques. Moreover, all the codewords in the CSCC codebook have exactly the same empirical distribution on a subblock level, enabling each codeword to be equally good for detection even in the worst-case scenario. We also provide a partial converse that matches our inner bound, demonstrating that our achievable codes are optimal within a certain class of typical codes. While our converse is limited to a certain class of codes, the main implication of the converse result is that our SCS detection strategy, detailed in Sec.~\ref{sec:main_results}, is indeed the asymptotically optimal detection rule for our codes. Finally, we present some example tradeoff regions for binary, scalar Gaussian, and multiple-input multiple-output (MIMO) Gaussian channels and discuss their implications. Particularly in the MIMO case, we demonstrate an intriguing tradeoff between sensing gains and multiplexing gains that add another degree of freedom in the tradeoff between information rate and QCD.

The remainder of the paper is organized as follows. In the next section, we provide some preliminaries for QCD. Then, in Sec.~\ref{sec:problem_statement}, we formally give the problem statement. In Sec.~\ref{sec:main_results}, we state our main results regarding the fundamental tradeoff for communication and QCD. In Sec.~\ref{sec:gaussian_channel}, we address some selective examples. Finally, in Sec.~\ref{sec:discussion}, we conclude the paper with discussions.

\textit{Notation:} We closely follow the notation in~\cite{El-Gamal--Kim2011}. Calligraphic letters denote alphabet spaces, and uppercase and lowercase letters respectively denote random variables and their realizations. For a sequence $x^n$ on $\Xc^n$, the type or composition of $x^n$ is defined to be $\pi(x | x^n) := \big| \{ i \colon x_i = x \} \big| /n$ for $x \in \Xc$. Then, $\Pc_n(\Xc)$ denotes the set of all types of $x^n$, and $\Tc_{p_X}^n$ with $p_X \in \Pc_n(\Xc)$ denotes the set of $x^n$ of type $p_X$. We also define $[1{:}n] = \{1,\ldots,n\}$ and $(x)^+=\max(0,x)$. The notation for mutual information $I(X;Y)$ can also be written as $I(p_X, p_{Y|X})$ to clarify its dependency. All logarithms are natural logarithms; hence, the unit of all information-theoretic quantities is nats.

\section{Preliminary on QCD} \label{sec:prelim}

In this section, we briefly summarize the standalone QCD problem. We refer to~\cite{Veeravalli--Banerjee2014} for a more detailed introduction to the topic.

Consider a sequence of observations $Y_1, Y_2, \ldots$ where the observations are initially i.i.d. with some distribution $p^{(0)}(y):=p(y|s=0)$, but at some unknown change point $\nu$, the distribution changes to $p^{(1)}(y):=p(y|s=1) \ne p^{(0)}(y)$. In other words,
\begin{align*}
	Y_i &\stackrel{\text{i.i.d.}}{\sim} p^{(0)} ~~~ \text{if } i < \nu, \nn\\
	Y_i &\stackrel{\text{i.i.d.}}{\sim} p^{(1)} ~~~ \text{if } i \ge \nu.
\end{align*}
The goal is to detect the change time $\nu$ as quickly as possible in an online manner but also to minimize the probability of declaring the change before $\nu$, i.e., the probability of false alarm.

Before discussing the QCD, let us formally define the stopping rule as follows.
\begin{definition}[stopping rule] \label{def:stopping_rule}
    A random variable $N \in \mathbb{N}$ is said to be a stopping rule with respect to a sequence of random variables $Y_1, Y_2, \ldots$ if $\{N = i\} \in \sigma(Y_1, \ldots, Y_i)$, where $\sigma(Y_1, \ldots, Y_i)$ is the $\sigma$-algebra generated by $(Y_1, \ldots, Y_i)$.
\end{definition}
In words, the stopping rule $N$ is a random variable such that the event $\{N = i\}$ is measurable with respect to the given information up to that time, i.e., the event $\{N=i\}$ is only a (possibly stochastic) function of $Y_1, \ldots, Y_i$.

Let $N$ be a stopping rule with respect to a sequence of observations $Y_1, Y_2, \ldots$. 
If we only consider detection delay, the problem becomes degenerate since to minimize detection delay $(N-\nu)^+$, it is best to declare the change point as early as possible, i.e., $N=1$ always, which will always result in a false alarm event unless $\nu = 1$. To penalize this, we measure the false alarm performance in terms of the mean time to a false alarm, or equivalently, its reciprocal called the false alarm rate (\FAR) given by
\begin{align}
	\FAR(N) = \frac{1}{\E_{\infty}(N)}, \label{eq:far}
\end{align}
where $\E_{\infty}$ assumes the probability distribution when the change occurs at $\nu = \infty$. For the objective function on detection delay, Lorden~\cite{Lorden1971} considered the worst case (over all possible change point $\nu \ge 1$) average detection delay conditioned on the worst realization $y^{\nu-1}$ (\WADD) as
\begin{align}
	\WADD(N) = \sup_{\nu \ge 1} \esssup_{Y^{\nu-1}} \E_\nu ( (N-\nu)^+ | Y^{\nu-1} ),\label{eq:vanilla_WADD}
\end{align}
where $\E_{\nu}$ assumes the probability distribution when the change occurs at $\nu\ge 1$.
With both measures at hand, Lorden's formulation characterizes the smallest $\WADD(N)$ subject to a {\FAR} constraint $\alpha > 0$, i.e.,
\begin{align}
	\inf_{N: \FAR(N) \le \alpha} \WADD(N). \label{eq:Lorden_formulation}
\end{align}

For QCD problems, the widely used stopping rules are based on the CuSum statistic, originated by Page~\cite{Page1954}, and its variations. The CuSum stopping rule with threshold $b$ is defined by
\begin{align}
	N_\CS = \inf \{ i \ge 1: W_i \ge b \},
\end{align}
where $W_i$ is the CuSum statistic defined by
\begin{align*}
	W_i = \max_{1 \le k \le i+1} \sum_{j=k}^{i} \log \frac{p^{(1)}(Y_j)}{p^{(0)}(Y_j)},
\end{align*}
where the right side is $0$ by convention if $k = i+1$.
It can also be written in a computationally efficient recursive form
\begin{align}
	 W_0 = 0, ~~ W_{i} = \left( W_{i-1} +\log \frac{p^{(1)}(Y_{i})}{p^{(0)}(Y_{i})} \right)^+.
\end{align}
Page's CuSum stopping rule $N_\CS$ was later shown by Lorden~\cite{Lorden1971} that it is indeed asymptotically optimal for Lorden's formulation~\eqref{eq:Lorden_formulation} as $\alpha \to 0$. Specifically, when $\alpha \to 0$,
\begin{align*}
    \inf_{N: \FAR(N) \le \alpha} \WADD (N) \sim \WADD(N_\CS) \sim \frac{| \log \alpha | }{ D(p^{(1)} \| p^{(0)}) },
\end{align*}
where $f \sim g$ denotes that $f/g \to 1$ as $\alpha \to 0$. Later, Moustakides~\cite{Moustakides1986} further showed that the CuSum algorithm is exactly optimal under Lorden's formulation for all $\alpha > 0$.

\section{Problem Statement} \label{sec:problem_statement}
In this section, we give a formal definition of our problem statement. 

Consider a state-dependent memoryless broadcast channel $p_{Y,\Yt|X,S}(y,\yt|x, s)$, shown in Fig.~\ref{fig:model1}, which can be decomposed into two conditionally independent components given by
\begin{align}
	p(y^n,\yt^n|x^n, s^n) = \prod_{i=1}^np(y_i|x_i,s_i)p(\yt_i|x_i)\label{eq:dmc}
\end{align}
where $S$ is the state, $X$ is the channel input, and $Y, \Yt$ are the outputs of the quickest change detection (QCD) channel and the state-independent communication channel, respectively. This model is widely studied in literature, e.g., \cite{Joudeh--Willems2022, Chang--Wang--Erdogan--Bloch--2023, hua2023mimo, liu2021cramer, Wu--Joudeh2022}.

Let $s \in \mathcal{S}=\{0,1\}$ where we note a base symbol by $0$. The sequence $s^n$ is a deterministic sequence
\begin{align}
	s^n = s_1, \ldots, s_{\nu-1}, s_{\nu}, \ldots, s_{n}\label{eq:state_seq}
\end{align}
where $\nu$ is a predetermined unknown change point from $s=0$ to $s=1$, i.e., 
\begin{align*}
	s_i = \begin{cases}
		0 & \text{ for } i < \nu, \\
		1 & \text{ for } i \ge \nu.
	\end{cases}
\end{align*}

In our problem setup, a code serves two distinct roles. First, it functions as a conventional channel code, used to communicate messages reliably over the noisy channel $p_{\Yt|X}$. 
Additionally, the transmitted code also acts as a sensing signal, through which we wish to detect the unknown change point $\nu$ over $p_{Y|X,S}$. Overall, a code for the problem setup is required to simultaneously transmit information and produce distinctive observations for QCD over the broadcast channel.

Formally, a code for joint communication and QCD has the following components. For a given block length $n$, an $(2^{nR}, n)$ code consists of
\begin{itemize}
	\item a message set $m \in [1:2^{nR}]$,
	\item an encoder that assigns a codeword $x^n(m)$ to each message $m\in[1:2^{nR}]$,
	\item a decoder that assigns a message estimate $\hat{m}(\yt^n)$ to each observation sequence $\yt^n$, and
    \item a stopping rule $N$ such that the event $\{N=i\}$ is measurable with respect to $Y^i$ and $x^n(m)$. If the stopping rule is not activated until the end of the codeword, we define $N = n+1$. 
\end{itemize}
A codebook is defined as $\Cc^{(n)} = \{x^n(m): m\in[1:2^{nR}]\}$. We emphasize that $N$ is dependent on $x^n(m)$, however, we omit this dependency in our notation for visual clarity.

The performance metrics for the problem are given as follows. For the communication component, the performance metric is the maximum probability of error denoted by
\begin{align*}
    \pen = \max_{m \in [1:2^{nR}]} \P ( \hat{M} \neq m | M=m ).
\end{align*}
We say that a rate $R \in \mathbb{R}_+$ is achievable if there exists a sequence of codes such that $\pen \to 0$ as $n \to \infty$.

Next, for the QCD metrics, we extend Lorden's formulation~\cite{Lorden1971} to our setting. To this end, fix a sequence of codebooks $\Ccn$. We first define the false alarm rate (\FAR) for the sequence as 
\begin{align}
	\FAR(N) &=  \limsup_{n\to\infty}\max_{x^n\in\Ccn}\frac{1}{\E_{\infty}( N | x^n )}. \label{eq:cond_FAR_max}
\end{align}
Similar to \eqref{eq:vanilla_WADD}, we define the {\WADD} objective for our setting, i.e., worst-case average detection delay over the worst possible realizations as
\begin{align}
	&\WADD(N) \nn \\
    &= \sup_{\nu \ge 1 }\limsup_{n\to\infty}\max_{x^n\in\Ccn}\esssup_{Y^{\nu-1}}\E_{\nu}((N-\nu+1)^+|x^n, Y^{\nu-1}).\label{eq:ps_wadd_max}
\end{align}
Then, we say that {\WADD} `$\Ds$' at {\FAR} constraint `$\alpha$' is achievable if there exists a sequence of codebooks $\mathcal{C}^{(n)}$ and stopping rule $N$ such that 
\begin{align*}
    \inf_{N, \{\Ccn\}: \FAR(N) \le \alpha} \WADD(N) \le \Ds.
\end{align*}

The ultimate goal of the joint communication and QCD problem is to understand the fundamental tradeoff between $R$ and $\Ds$ for some fixed {\FAR} constraint $\alpha$, i.e., we aim to answer ``What is the maximum information rate $R$ and minimum delay $\Ds$ that a code can simultaneously achieve?''. In particular to this work, we are interested in an asymptotic regime of QCD where $\alpha \to 0$. To better characterize the tradeoff in this regime, we define 
\begin{align}
	\Delta := \lim_{\alpha \to 0} \frac{|\log \alpha|}{\Ds}\label{eq:Delta}
\end{align}
and say that $\Delta$ is achievable if there exists a pair $(\alpha, \Ds)$ that asymptotically attains $\Delta$ as $\alpha \to 0$. The reason for an additional definition $\Delta$ will be apparent when we introduce our achievable region in Thm.~\ref{thm:theorem1}, in which $\Delta$ is simply characterized in the form of a KL divergence. Moreover, $\Delta$ represents the asymptotic ``slope'' of $\Ds$ as a function of $\alpha$ or can be thought of as the speed of sensing. Then, the goal of this work can be condensed to characterizing the set of achievable $(R, \Delta)$ pairs, namely, the $R$-$\Delta$ region
\begin{align}
	\mathscr{R} &= \text{closure} \{ (R, \Delta ): \text{$R$ and $\Delta$ are achievable} \}. \label{eq:region}
\end{align}
Equivalently, we can express $\mathscr{R}$ using the $R$-$\Delta$ function
\begin{align}
    R(\Delta) := \sup \{ R: (R, \Delta') \text{ is achievable for some } \Delta' \le \Delta \}\label{eq:r_delta_func}
\end{align} and the $\Delta$-$R$ function
\begin{align}
    \Delta(R) := \sup \{ \Delta: (R', \Delta) \text{ is achievable for some } R' \le R \} \label{eq:delta_r_func}
\end{align} as
\begin{align*}
	\mathscr{R} &= \{ (R, \Delta): R \le R(\Delta), \Delta \in \mathbb{R}_+ \}, \\
	&= \{ (R, \Delta): R \in \mathbb{R}_+, \Delta \le \Delta(R) \}.
\end{align*}

To give some perspective to the problem, we can immediately identify two extreme operating points. On one extreme, we can design codes that maximize the communication rate while neglecting QCD. By Shannon's capacity-achieving codes, the operating point $(C, 0)$ where $C = \max_{p_X}I(X;\Yt)$ is an achievable pair. On the other extreme, we can design codes for pure QCD detection which results in the operation point $(0, \Delta^\star)$~\cite{Veeravalli--Banerjee2014}, where 
\begin{align}
	\Delta^\star := \max_{x\in\Xc} D(p^{(1)}\| p^{(0)}|x),\label{eq:delta_star}
\end{align}
that is, the optimal $\Delta$ when only QCD is considered.
Extending on the two extreme points, we can alternate transmission between capacity-achieving codes and the optimal QCD input symbol via timesharing. This results in half of the time sending information and half of the time sending a QCD signal in an alternating fashion. Since we are communicating information only half of the time, $C/2$ is achieved for communication, and since it takes twice as long for QCD, $\Delta^\star/2$ is achieved for detection. Generalizing the timesharing strategy for a timesharing fraction $\lambda\in[0,1]$, $(\lambda C, (1-\lambda)\Delta^\star)$ is achievable. In addition, the point $(C,\Delta^\star)$ will serve as a trivial outer bound of $\mathscr{R}$. 

\section{Main Results}\label{sec:main_results}
In this section, we state our main results and discuss some implications. 
\begin{theorem}\label{thm:theorem1}
	Let $\mathscr{R}_{\inn}$ be the set of rate-delay pairs $(R, \Delta) \in \mathbb{R}_+^2$ such that 
	\begin{align}
		R &< I(X;\Yt), \label{eq:thm1_rate}\\
        \Delta &< D(p^{(1)} \| p^{(0)} | p_X), \label{eq:thm1_delta}
	\end{align}
	for some $p_X$, where $p^{(s)} := p_{Y|X,s}$. Then, $\mathscr{R}_{\inn} \subset \mathscr{R}$. In other words, any point in $\mathscr{R}_{\inn}$ is achievable.
\end{theorem}
In the following, after some remarks, we provide an overview of the achievability strategy. The detailed proof of Thm.~\ref{thm:theorem1} is deferred to App.~\ref{sec:proof_thm1}.

\begin{remark}
    The region $\mathscr{R}_\inn$ is convex since the mutual information and the KL divergence are concave and linear in $p_X$, respectively. Thus, including timesharing does not expand the $\mathscr{R}_\inn$ region.
\end{remark}

\begin{remark} The region $\mathscr{R}_\inn$ includes two extreme points, $(C, 0)$ and $(0, \Delta^\star)$, where 
    \begin{align*}
        C = \max_{p_X} I(X;\Yt), \quad \Delta^\star = \max_{x\in\Xc} D(p^{(1)}\|p^{(0)}|x).
    \end{align*}
\end{remark}

\begin{remark}\label{rmk:BA_algorithm}
    By the concavity of the mutual information and viewing~\eqref{eq:thm1_delta} as a constraint on the distribution, i.e.,
    \begin{align}
    \max_{p_X: \E(c(X)) > \Delta } I(X;\Yt) \label{eq:BA_objective}
    \end{align} 
    where $c(x) = D(p^{(1)} \| p^{(0)}|X=x)$, we can easily apply the Blahut--Arimoto algorithm~\cite{Arimoto1972,Blahut1972} with input constraints given in~\cite[Sec.~IV]{Blahut1972} to evaluate the tradeoff region. The algorithm for our case is summarized in App.~\ref{app:BA}.
\end{remark}

For the achievability proof, we use the constant subblock-composition codes (CSCCs) originally developed in~\cite{Tandon--Motani--Varshney2016}.
The role of using CSCCs is in line with the motivation of using constant composition codes as in~\cite{Zhang--Vedantam--Mitra2011, Joudeh--Willems2022, Chang--Wang--Erdogan--Bloch--2023}, i.e., we wish to provide some worst-case performance guarantee for QCD. However, in our setting, we use CSCCs instead of conventional constant composition codes to guarantee an additional property that any {\em subsequence} of the codes has approximately a constant composition. To this end, we wish to design codewords that have a universal property such that any large enough interval of the codeword belongs to a specific type. For any codeword in the CSCC and any given change point $\nu$, a large enough interval with a ``small'' shift to the start of a subblock guarantees that the interval belongs to a specific type. Since this is a universal property for all codewords in the CSCC codebook, it provides the basis for our worst-case analysis that follows while the small shift will be negligible in our analysis.

The codewords in the CSCC has an identical composition of symbols in each subblock which provides consistent statistical properties between the codewords. In particular, the statistical properties in the pre-change interval after some permutation are identical between the codewords. The same property holds for the post-change codeword statistics. This type of design ensures that there are no particularly ``bad'' codewords which helps for our worst-case delay metric.

On the other hand, this is not possible with Shannon's random code nor CCC since a certain codeword could have a nonidentical composition of symbols in earlier and later parts of the codeword, leading to nonuniform QCD performance depending on the codeword and change point.

We now explain our code construction. Let $\Pc_L(\Xc)$ denote the set of all compositions of $x^L$, and let $\Tc_{p_X}^L$ be the set of sequences in $\Xc^L$ with composition $p_X$. Fix $p_X \in \Pc_L(\Xc)$, and let $R$ be the rate of the CSCC. Further, let $n$ be the codeword length that is composed of $k$ subblocks of length $L$, i.e., $n=kL$. Then, 
\begin{align}
    x^n = \xv^k = (\xv_1,\ldots, \xv_k),
\end{align}
where $\xv_j = x_{(j-1)L+1}^{jL}$, $j\in[1:k]$. Thus, we can view $x^n$ as a $k$-length sequence with $L$-length vector symbols. Then, the codebook is generated as follows.

{\noindent \bf Codebook generation:} Fix $p_X$, and randomly and independently generate $2^{nR}$ sequences $\xv^k(m)$ such that each vector of length $L$ is i.i.d.~uniformly distributed over $\Tc_{p_X}^L$. The codebook is defined as $\Cc^{(n)} = \{\xv^k(m): m\in[1:2^{nR}]\}$.

As shown in~\cite{Tandon--Motani--Varshney2016}, by letting $k\to\infty$, the following rate is achievable using CSCC:
\begin{align}\label{eq:mi_bound_L}
    I(p_X, p_{\Yt|X}) -r(L, p_X) \le R \le I(p_X, p_{\Yt|X}),
\end{align}
where the term $r(L, p_X)$ vanishes with $L$, which will be specified in App.~\ref{sec:proof_thm1}. Then, by letting $L\to\infty$, rate $R < I(X;\Yt)$ is achievable. Incorporating a deliberate detour using CSCC to show that $R<I(p_X, p_{\Yt|X}) = I(X;\Yt)$ is achievable serves as a crucial component in our QCD analysis, which follows below.

For the QCD analysis, aligning with the vector-symbol interpretation of $x^n$, let us view the QCD channel as a discrete memoryless channel over $\xv$, i.e., $p^{(s)}(\yv^k|\xv^k) = \prod_{j=1}^kp^{(s)}(\yv_j|\xv_j)$, where $\yv_j=y_{(j-1)L+1}^{jL}$ and $p^{(s)}(\yv_j|\xv_j)=\prod_{i=1}^L p^{(s)}_{Y|X}(y_{(j-1)L+i})|x_{(j-1)L+i})$, $s=0,1$. Note that the observations $\yv_1, \yv_2, \ldots, \yv_k$ are produced from subblocks of the same composition and can be made i.i.d.~within the pre-change and post-change regimes by proper permutation.

Next, we propose to use an adapted CuSum test, which we call a subblock-CuSum (SCS) test. Instead of the standard CuSum statistic that updates the statistic whenever it sees a new observation, the subblock-CuSum test updates it only at the end of subblocks. Formally, letting $W_0 = 0$, we construct the SCS statistic as follows:
\begin{align}
    W_{i} &= \begin{cases}
        \left( W_{i-1} + \log \frac{ p^{(1)}_{\Yv|\Xv}(\Yv_j |\Xv_j)}{ p^{(0)}_{\Yv|\Xv}(\Yv_j |\Xv_j) }  \right)^+ & \text{if } i=jL, j\in\mathbb{N}\\
        W_{i-1} & \text{otherwise}. \label{eq:SCS}
    \end{cases}
\end{align}
When $W_i \ge b$, we stop updating and declare that a change occurred, i.e.,
\begin{align}
    N_{\SCS} := \inf \{ i \ge 1: W_i \ge b \}. \label{eq:NSCS}
\end{align}
With the combination of CSCCs and SCS, for some $\alpha>0$, we formally show in App.~\ref{sec:proof_thm1} that as $\alpha \to 0$, it asymptotically holds that
\begin{align*}
    \E_\infty (N_{\SCS}) \ge \frac{1}{\alpha}
\end{align*}
and
\begin{align*}
    \WADD(N_{\SCS}) \le \frac{ |\log \alpha| }{ D(p^{(1)} \| p^{(0)} | p_X) } (1+o(1)) =: \Ds.
\end{align*}
Then, our CSCC code achieves $\Delta$ such that
\begin{align*}
    \Delta = \lim_{\alpha \to 0} \frac{ |\log \alpha| }{\Ds} = D( p^{(1)} \| p^{(0)} | p_X).
\end{align*}

Next, we discuss two conflicting effects of the subblock length $L$. First, achieving the channel capacity with CSCCs, as stated in \eqref{eq:mi_bound_L}, generally requires a close approximation of the capacity-achieving distribution $p_X$, which in turn requires a large $L$ for information-theoretic completeness. However, in the case of QCD, while the impact of a large $L$ does not appear in our asymptotic characterization, a longer subblock length introduces a larger delay because the SCS statistics are only updated every subblock. Therefore, aside from the need to approximate a target $p_X$, shorter $L$ is desirable to minimize the QCD delay. We also remark that there exists examples where finite $L$ suffices to attain the capacity of the communication channel, e.g. a binary symmetric channel.

There are some extensions of Thm.~\ref{thm:theorem1} that are due. First, consider the case where the state alphabet is $|\Sc|>2$. The definition of {\WADD} for this case is given by,
\begin{align}
	\WADD(N) := \bar{\E}_{\nu,s}((N-\nu+1)^+|x^n, Y^{\nu-1}).\label{eq:ps_wadd_multis}
\end{align}
where 
\begin{align*}
\bar{\E}_{\nu,s}(\cdot):=\max_{s\in\Sc\backslash\{0\}} \sup_{\nu \ge 1} \limsup_{n \to \infty} \max_{x^n \in \mathcal{C}^{(n)}} \esssup_{Y^{\nu-1}} \E_{\nu,s}(\cdot).
\end{align*}
Then, we can perform $|\Sc|-1$ SCS tests in parallel and pick the quickest change point estimate among the tests. This extension leads to the following corollary.

\begin{corollary}\label{cor:multi-state}
    For the case $|\Sc|>2$, the set of rate-delay pairs $(R, \Delta) \in \mathbb{R}_+^2$ such that 
	\begin{align}
		R &< I(X;\Yt),\\ 
        \Delta &< \min_{s\in\Sc\backslash\{0\}}D(p_{Y|X,s} \| p^{(0)} | p_X), \label{eq:cor_delta}
	\end{align}
	for some $p_X$ is achievable.
\end{corollary}

\begin{remark}
    By the standard quantization argument, e.g., \cite[Chap.~3.4.1]{El-Gamal--Kim2011}, constant subblock-composition codes can be extended for Gaussian channels,  which then allows the rate region $\mathscr{R}_\inn$ to be applied to the Gaussian case. We provide some discussions on some Gaussian settings in Sec.~\ref{sec:gaussian_channel}.
\end{remark}

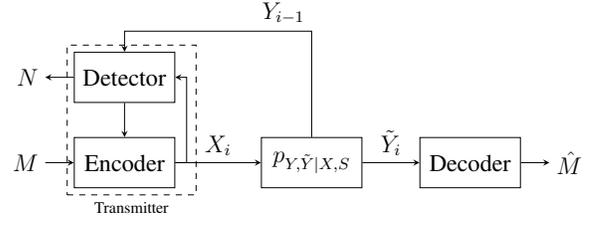
\begin{figure}[!t]
	\vspace{-0em}
	\centering
\resizebox{!}{8.5em}{\begin{tikzpicture}[font=\large, node distance=.6cm and 1cm, start chain]
     \tikzstyle{rect}=[draw=black, 
                   rectangle, 
                   text opacity=1,
                   minimum width=50pt, 
                   minimum height = 25pt, 
                   align=center]
  \node[rect] (encoder) {Encoder};
  \node[rect, above=of encoder] (detector) {Detector};
  
  \node[rect, right=1.5cm of encoder] (chn) {$p_{Y,\tilde{Y}|X, S}$};
  \node[rect, right=of chn] (decoder) {Decoder};

    \draw[-stealth] (encoder.east) node (line1)[right]{}-- +(0.2,0) |- (detector.east);
  \node[fit=(encoder)(detector)(line1), label={[font=\footnotesize]below:Transmitter}, draw, dashed] (transmitter) {};

\draw[-stealth] (encoder.east) -- node[above] {$X_i$} (chn.west);
\draw[-stealth] (detector.south) -- (encoder.north);
\draw[-stealth] (chn.north)  |- node[above left] {${Y}_{i-1}$} (0,2.3cm) --  (detector.north);
\draw[stealth-] (encoder.west) --+ (-5mm,0) node[left] {$M$};
\draw[-stealth] (detector.west) --+ (-5mm,0) node[left] {$N$};
\draw[-stealth] (chn.east) -- node[above] {$\Yt_i$}(decoder.west);
\draw[-stealth] (decoder.east) --+ (5mm,0) node[right] {$\hat{M}$};
\end{tikzpicture}}
	\caption{Monostatic model for ISAC. The encoder and QCD detector are the same entity which has access to some feedback.}
	\label{fig:monostatic_model}
    \vspace{-0.15in}
\end{figure}

Next, we discuss some applications to the monostatic model given in Fig.~\ref{fig:monostatic_model} where $p(y,\yt|x,s)=p(y|x, s)p(\yt|x)$. The only difference from the problem statement of the bistatic case is that the encoder is defined by a mapping that assigns a codeword $x_i(m, y^{i-1})$, $i\in[1:n]$ to each message $m\in[1:2^{nR}]$ as well as a causal observation sequence $y^{i-1}$. For this case, Thm.~\ref{thm:theorem1} remains an inner bound since we can disregard the observation $y^{i-1}$ in the codebook design and use the same codebook and QCD detector defined for the bistatic case. In general, the encoder structure for the bistatic case could be improved by incorporating the feedback information $y^{i-1}$ at the encoder, i.e., a code for the bistatic model can be thought of as a baseline code for the monostatic model that does not use adaptive strategy. Some discussions on extensions to the bistatic case are given in Sec.~\ref{sec:discussion}.

In the following, we show that Thm.~\ref{thm:theorem1} is indeed a tight evaluation of CSCCs for our joint communication and QCD setting, i.e., no other QCD detectors can be asymptotically better than the SCS detector. The converse result is shown on a slightly wider class of codes defined below.

\begin{definition}[sliding-window typical codes]
	A code $\mathcal{C}^{(n)}$ is \textit{sliding-window typical} with $p_X$ if for any $\epsilon > 0$, there exists a window size $L_0(\epsilon)$ such that for every $x^n \in \mathcal{C}^{(n)}$ and $i \in \mathbb{N}$ with $i \le n - L_0 +1$, it holds that
	\begin{align}
		\left| \pi(a|x_i^{i + L_0 - 1}) - p_X(a)  \right| \le \epsilon p_X(a) ~~~ \text{for all $a \in \mathcal{X}$ }, \label{eq:sliding_typical}
	\end{align}
 where $\pi(a|x_i^{i + L_0 - 1})$ is the type of the length $L_0$ sequence $x_i^{i + L_0 - 1}$.
Also, a sequence of codes $\Ccn$ is sliding-window typical with $p_X$ if every $\Ccn$ is sliding-window typical with $p_X$.
\end{definition}
\begin{remark}
    We note that the proposed CSCC is included in the class of sliding-window typical codes as $n\to\infty$. To see this, we can choose a large enough $L_0$ such that it includes enough fully enclosed subblocks with type $p_X^{(L)}$ such that it satisfies the $\epsilon$-typicality condition in~\eqref{eq:sliding_typical} for $P_X=P_X^{(L)}$. For arbitrary $P_X$, we take $L\to\infty$. 
\end{remark}

Then, we have the following lower bound on {\WADD} for the class of sliding-window typical codebooks.
\begin{theorem}\label{thm:theorem2}
	For a sequence of codebooks $\Ccn$ that is sliding-window typical with $p_X$ and any stopping time $N$, the QCD delay {\em \WADD} with $\text{\em \FAR}(N) \le \alpha$ constraint is bounded by
	\begin{align*}
		\text{\em \WADD}(N) \ge |\log \alpha| \left( \frac{1}{ D( p^{(1)} \| p^{(0)} | p_X ) } + o(1) \right)
	\end{align*}
 as $\alpha \to 0$.
\end{theorem}
\begin{IEEEproof}
The proof is deferred to App.~\ref{sec:proof_thm2}.
\end{IEEEproof}

The above theorem states that no matter what stopping rule is used, {\WADD} cannot be asymptotically smaller than $\frac{ |\log \alpha| }{ D( p^{(1)} \| p^{(0)} | p_X ) }$. As the lower bound coincides with the delay characterization in Thm.~\ref{thm:theorem1}, our SCS stopping rule is asymptotically optimal for CSCCs, or more generally, the class of sliding-window typical codes. 

\begin{theorem}\label{thm:theorem3}
	For the class of sliding-window typical codes, $\mathscr{R}_\inn$ in Thm.~\ref{thm:theorem1} is tight, i.e., on the space of sliding-window typical codes $\mathscr{R}_\inn = \mathscr{R}_{\pi}$, where $\mathscr{R}_{\pi}$ is the rate-delay region for sliding-window typical codes.
\end{theorem}
\begin{IEEEproof}
The proof is deferred to App.~\ref{sec:proof_thm3}.
\end{IEEEproof}

Unlike standard converse proofs that consider {\em any} codes, the above converse is limited to the class of sliding-window typical codes. However, the above converse has an important implication; in particular, it shows that our SCS detection algorithm is in fact asymptotically optimal for the CSCC codes. This is in a similar spirit to the results in~\cite{Bandemer--El-Gamal--Kim2015} where it shows a converse only for the class of i.i.d.~randomly generated codes and proves that the proposed decoder is optimal, i.e., equivalent to maximum likelihood (ML) decoders. In our setting, showing the optimality of the detector is even more prominent since we do not have an evident optimal detector (c.f.~ML for decoding), and defining a sequential detector is an essential part of the problem for QCD.

\section{Evaluation Examples}\label{sec:gaussian_channel}
In this section, we demonstrate how our results can be applied through some selection of example cases.

\subsection{Binary channels}\label{sec:BSC_example}
\begin{figure}[!tb]
	\centering
	\begin{subfigure}[b]{0.2\textwidth}
		\centering
            \resizebox{!}{7em}{\begin{tikzpicture}[node distance=.6cm and 1cm, start chain]

  \node (input0) {$0$};
  \node[below=1.5cm of input0] (input1) {$1$};
  \node[right=2.5cm of input0] (output0) {$0$};
  \node[right=2.5cm of input1] (output1) {$1$};

\draw[-] (input0.east) -- node[above] {} (output0.west);
\draw[-] (input1.east) -- node[below] {$1-\epsilon_s$} (output1.west);
\draw[-] (input1.east) -- node[below, xshift=0.1cm] {$\epsilon_s$} (output0.west);
\end{tikzpicture}}
		\caption{$p^{(s)}_{Y|X}(y|x)$}
		\label{fig:bin_ps0}
	\end{subfigure}
 \begin{subfigure}[b]{0.2\textwidth}
		\centering
		\resizebox{!}{8em}{\begin{tikzpicture}[node distance=.6cm and 1cm, start chain]

  \node (input0) {$0$};
  \node[below=1.5cm of input0] (input1) {$1$};
  \node[right=2.5cm of input0] (output0) {$0$};
  \node[right=2.5cm of input1] (output1) {$1$};

\draw[-] (input0.east) -- node[above] {$1-\epsilon$} (output0.west);
\draw[-] (input1.east) -- node[below] {$1-\epsilon_{}$} (output1.west);
\draw[-] (input0.east) -- node[above, xshift=0.5cm] {$\epsilon$} (output1.west);
\draw[-] (input1.east) -- node[below, xshift=0.5cm] {$\epsilon$} (output0.west);
\end{tikzpicture}}
		\caption{$p_{\Yt|X}(\yt|x)$}
		\label{fig:bin_pcom}
	\end{subfigure}
	\caption{Channels considered in the binary example.}
	\label{fig:binary_ex}
\end{figure}
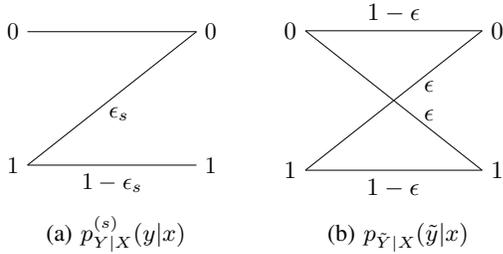

\begin{figure}[!tb]
	\vspace{-0em}
	\centering
\resizebox{!}{13.5em}{
%
%
\begin{tikzpicture}

\begin{axis}[%
width=3.5in,
height=2.5in,
scale only axis,
separate axis lines,
every outer x axis line/.append style={white!15!black},
every x tick label/.append style={font=\Large\color{white!15!black}},
xmin=0,
xmax=0.6,
xlabel={\Large $\Delta$},
every outer y axis line/.append style={white!15!black},
every y tick label/.append style={font=\Large\color{white!15!black}},
ymin=0,
ymax=0.09,
ylabel={\Large Rate [nats/c.u.]},
legend style={draw=white!15!black,fill=white,legend cell align=left}
]
\addplot [color=blue,solid,line width=2.0pt,forget plot]
  table[row sep=crcr]{%
0	0.0822828785050518\\
0.255412811882995	0.0822828785050518\\
0.257992739275753	0.0822747160504383\\
0.26057266666851	0.0822502284200882\\
0.263152594061268	0.0822094148144191\\
0.265732521454025	0.0821522739006209\\
0.268312448846783	0.0820788038123926\\
0.270892376239541	0.081989002149578\\
0.273472303632298	0.0818828659776943\\
0.276052231025056	0.0817603918273568\\
0.278632158417813	0.0816215756935971\\
0.281212085810571	0.0814664130350767\\
0.283792013203328	0.0812948987731944\\
0.286371940596086	0.0811070272910867\\
0.288951867988843	0.0809027924325207\\
0.291531795381601	0.0806821875006818\\
0.294111722774358	0.0804452052568499\\
0.296691650167116	0.08019183791897\\
0.299271577559873	0.0799220771601105\\
0.301851504952631	0.0796359141068131\\
0.304431432345388	0.0793333393373313\\
0.307011359738146	0.0790143428797562\\
0.309591287130904	0.0786789142100303\\
0.312171214523661	0.0783270422498458\\
0.314751141916419	0.0779587153644303\\
0.317331069309176	0.0775739213602146\\
0.319910996701934	0.0771726474823836\\
0.322490924094691	0.0767548804123089\\
0.325070851487449	0.0763206062648619\\
0.327650778880206	0.0758698105856046\\
0.330230706272964	0.0754024783478593\\
0.332810633665721	0.0749185939496525\\
0.335390561058479	0.0744181412105348\\
0.337970488451236	0.0739011033682717\\
0.340550415843994	0.0733674630754068\\
0.343130343236751	0.0728172023956933\\
0.345710270629509	0.0722503028003922\\
0.348290198022266	0.0716667451644381\\
0.350870125415024	0.0710665097624645\\
0.353450052807781	0.0704495762646942\\
0.356029980200539	0.0698159237326855\\
0.358609907593296	0.0691655306149378\\
0.361189834986054	0.0684983747423481\\
0.363769762378812	0.0678144333235232\\
0.366349689771569	0.0671136829399378\\
0.368929617164327	0.0663960995409408\\
0.371509544557084	0.0656616584386064\\
0.374089471949842	0.064910334302423\\
0.376669399342599	0.064142101153824\\
0.379249326735357	0.0633569323605506\\
0.381829254128114	0.0625548006308475\\
0.384409181520872	0.0617356780074868\\
0.386989108913629	0.0608995358616177\\
0.389569036306387	0.0600463448864369\\
0.392148963699144	0.0591760750906761\\
0.394728891091902	0.0582886957919054\\
0.397308818484659	0.0573841756096451\\
0.399888745877417	0.0564624824582846\\
0.402468673270174	0.0555235835398025\\
0.405048600662932	0.0545674453362832\\
0.40762852805569	0.0535940336022295\\
0.410208455448447	0.0526033133566588\\
0.412788382841205	0.051595248874987\\
0.415368310233962	0.0505698036806882\\
0.41794823762672	0.0495269405367271\\
0.420528165019477	0.0484666214367618\\
0.423108092412235	0.0473888075961054\\
0.425688019804992	0.046293459442447\\
0.42826794719775	0.0451805366063208\\
0.430847874590507	0.0440499979113202\\
0.433427801983265	0.0429018013640506\\
0.436007729376022	0.0417359041438111\\
0.43858765676878	0.0405522625920028\\
0.441167584161537	0.0393508322012536\\
0.443747511554295	0.0381315676042505\\
0.446327438947052	0.0368944225622782\\
0.44890736633981	0.0356393499534468\\
0.451487293732568	0.0343663017606107\\
0.454067221125325	0.0330752290589607\\
0.456647148518083	0.0317660820032862\\
0.45922707591084	0.0304388098148983\\
0.461807003303598	0.0290933607681995\\
0.464386930696355	0.0277296821768968\\
0.466966858089113	0.0263477203798431\\
0.46954678548187	0.0249474207264978\\
0.472126712874628	0.0235287275619959\\
0.474706640267385	0.0220915842118151\\
0.477286567660143	0.0206359329660258\\
0.4798664950529	0.0191617150631159\\
0.482446422445658	0.0176688706733744\\
0.485026349838415	0.0161573388818231\\
0.487606277231173	0.0146270576706794\\
0.49018620462393	0.01307796390134\\
0.492766132016688	0.0115099932958669\\
0.495346059409446	0.00992308041796319\\
0.497925986802203	0.00831715865341975\\
0.500505914194961	0.00669216019002006\\
0.503085841587718	0.00504801599688076\\
0.505665768980476	0.00338465580321456\\
0.508245696373233	0.00170200807649457\\
0.510825623765991	0\\
};
\addplot [color=red,line width=2.0pt,only marks,mark=o,mark options={solid},forget plot]
  table[row sep=crcr]{%
0.255412811882995	0.0822828785050518\\
};
\addplot [color=red,solid,line width=2.0pt,forget plot]
  table[row sep=crcr]{%
0	0.0822828785050518\\
0.510825623765991	0\\
};
\end{axis}
\end{tikzpicture}%
}
	\caption{Plot for the binary channel example where $\epsilon = 0.3$, $\epsilon_0 = 0.1$, $\epsilon_1 = 0.5$. }
	\label{fig:model}
    \vspace{-0.15in}
\end{figure}
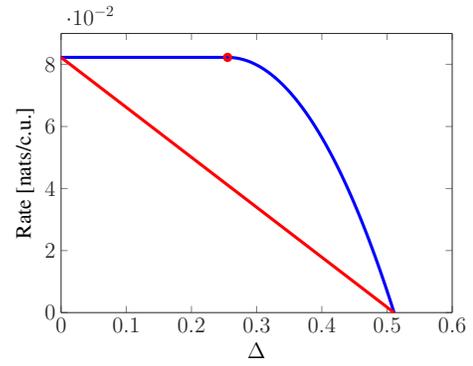
Consider the case in which the communication channel $p_{\Yt|X}$ is the standard binary symmetric channel with crossover probability $\epsilon$ and the QCD channel $p^{(s)}_{Y|X} = p_{Y|X, S}(y|x,s)$ is a binary $Z$-channel given by
\begin{align}
	p^{(0)}(0|0) &= 1, \quad p^{(0)}(0|1) = \epsilon_0 \\
	p^{(1)}(0|0) &= 1, \quad p^{(1)}(0|1) = \epsilon_1.
\end{align}
The example is contrived so that sending the symbol $x=0$ is useless for the QCD as it always outputs $Y=0$ symbol in pre- and post-change regimes. However, sending $x=0$ and $x=1$ equally likely maximizes the capacity of the communication channel, which implies there will be a tradeoff between communication and QCD. 

Figure~\ref{fig:model} depicts the tradeoff for our proposed strategy at $\epsilon = 0.3$, $\epsilon_0 = 0.1$, $\epsilon_1 = 0.5$. The linear line is a baseline tradeoff given by timesharing between pure communication and QCD. The capacity of the communication channel is given by $H_2(p_X)-H_2(\epsilon)$ with $H_2(\cdot)$ being the binary entropy function, which is attained by a uniform distribution $p_X$. On the other hand, the optimal QCD-only point is $\Delta^\star = D(p^{(1)}\| p^{(0)}|X=1)$, i.e., QCD with a constant input symbol $x=1$. The overall achievable region is given by the tradeoff $(H_2(p_X)-H_2(\e), D(p^{(1)}\| p^{(0)}|p_X))$ where 
\begin{align*}
    &D(p^{(1)}\|p^{(0)}|p_X)\\
    & = p_X(0)D(p^{(1)}\|p^{(0)}|X=0)+p_X(1)D(p^{(1)}\|p^{(0)}|X=1) \\
    & = p_X(1)\left(\epsilon_1\log\frac{\epsilon_1}{\epsilon_0}+(1-\epsilon_1)\log\frac{(1-\epsilon_1)}{(1-\epsilon_0)}\right).
\end{align*}
The circle dot point is the joint communication and QCD achievable point attained while maintaining capacity-achieving communication $(C, \Delta(C))$, i.e., when $p_X$ is uniform.

\subsection{Scalar Gaussian channels}\label{sec:Gaussian_example}
Consider the following Gaussian channels for QCD
\begin{alignat*}{2}
	Y_i &= x_i + Z_i \quad &\text{ for } i< \nu, \\
	Y_i &= hx_i + Z_i \quad &\text{ for } i\ge \nu,
\end{alignat*}
and the communication channel given by $\Yt_i = x_i + \Zt_{i}$, where $Z_i, \Zt_{i}$ are independent standard Gaussian noise. The channel input is subject to an average power constraint $\frac{1}{n}\sum_{i=1}^n x_i^2 \le P$.
This setting represents the case, for example, when a user is jointly communicating and sensing for some disruption in its path, which will alter the channel gain.

The capacity of the communication channel is $C = \frac{1}{2} \log (1+P)$, attained by Gaussian input with $\Nc(0, P)$. For the delay of QCD, note that when $X = x$, $D(p^{(1)} \|p^{(0)}|X=x) =  \frac{(h-1)^2}{2} x^2$. It in turn implies that $\Delta^\star$ in~\eqref{eq:delta_star} is
\begin{align*}
    \Delta^\star = \max_{ x : x^2 \le P } D(p^{(1)}\| p^{(0)}|x) = \frac{(h-1)^2}{2} P.
\end{align*}
On the other hand, due to the power constraint,
\begin{align*}
	\Delta = D(p^{(1)} \|p^{(0)}|p_X) = \int \frac{(h-1)^2}{2} x^2 p(x) dx \le \frac{(h-1)^2}{2} P,
\end{align*}
where the equality holds if the input is Gaussian distributed with $\Nc(0, P)$. Thus, by choosing $X \sim \Nc(0, P)$, the optimal communication rate and QCD delay are attained simultaneously, i.e., $(C, \Delta^\star)$ is achievable where
\begin{align*}
	(C,\Delta^\star) = \left(\frac{1}{2}\log(1+P),\, \frac{(h-1)^2}{2}P\right).
\end{align*}
Since the optimal QCD and the channel capacity are simultaneously achieved, $\mathscr{R}_\inn=\mathscr{R}$ for this case.

Next, consider the following Gaussian channels, where the sensing channel is given by
\begin{align*}
	Y_i &= x_i + Z^{(0)}_i \quad \text{ for } i\le \nu, \\
	Y_i &= x_i + Z^{(1)}_i \quad \text{ for } i> \nu,
\end{align*}
where $Z^{(s)}_i \sim \Nc(0, \sigma^2_s)$, $s=0,1,$ and the communication channel is the same as the previous example. This setting represents the case, for example, when a user is jointly communicating and sensing the existence of interference signals.

In this case, we can show that $\Delta^\star$ in~\eqref{eq:delta_star} is given by
\begin{align*}
    D(p_1 \|p_0 | p_X) = \frac{1}{2} \log\frac{\sigma^2_0}{\sigma^2_1} + \frac{\sigma^2_1}{2\sigma^2_0} - \frac{1}{2} = \Delta^\star,
\end{align*}
which is independent of $p_X$. Thus,
\begin{align*}
	(C, \Delta^\star) = \left(\frac{1}{2}\log(1+P), \frac{1}{2} \log\frac{\sigma^2_0}{\sigma^2_1} + \frac{\sigma^2_1}{2\sigma^2_0} - \frac{1}{2}\right)
\end{align*}
is attainable, which means that the operating point by the Gaussian input with power $P$ is optimal for both communication and QCD, i.e., $\mathscr{R}_\inn=\mathscr{R}$.

For both cases, no tradeoff between rate and delay is observed. Interestingly, it coincides with a prior conclusion for binary state detection under Gaussian noise \cite{Joudeh--Willems2022}.

\subsection{Example application to passive radar detection}
We provide a simplified example that illustrates how our results can be applied to a passive radar system~\cite{Poullin2005, Salah2013}. We consider a network with two base stations and a communication receiver, in which one of the base stations broadcasts an ISAC signal to the other base station and the communication receiver. The goal of the communication receiver is to recover the message from the ISAC signal, while the second base station serves as a passive radar detector having the message as side information via a wireline link. The goal of the radar receiver is to detect a target, which is present only after unknown time $\nu$, as soon as possible. We assume that the received signal has a direct path component, and additionally, a target deflected path component~\cite{Hack2013, Zheng--Lops--Eldar--Wang2019} is present after time $\nu$. Formally, the discretized baseband signal for the passive radar system is given by 
\begin{align*}
    \bar Y_i &= h_0 x_{i-\tau_d} + \bar Z_i &\text{for } i< \nu+\tau, \\
    \bar Y_i &= h_0 x_{i-\tau_d} +h e^{j2\pi f i} x_{i-\tau}+\bar Z_i &\text{for } i\ge \nu+\tau,
\end{align*}
where $\sum_i |x_i|^2\le nP$, $\bar Z_i$ is the complex additive white Gaussian noise (AWGN) with $\Cc\Nc(0,1)$, $\tau$ is the path delay (measured in samples), $h$ is the complex channel coefficient, and $f$ is the normalized Doppler frequency of the target deflected path. The parameters $\tau_d$, $h_0$ are defined similarly for the direct path. In general, parameters such as channels, Doppler shift, and delay are simultaneously estimated with target detection~\cite{Zaimbashi2017, Palmer2013, Liu--Li--Himed2014}. 
However, to narrow our scope to the QCD aspects, we assume that these parameters are perfectly estimated. A fundamental tradeoff between the probability of detection error and communication rate has been studied in a similar setup in~\cite{An--Li--Ng--Yuen2023}. 

Since the transmitted message is known at the QCD detector, we can subtract the direct path components from the received signal, which gives 
\begin{align*}
    \bar Y_i &= \bar Z_i &\text{for } i< \nu+\tau, \\
    \bar Y_i &= h e^{j2\pi f i} x_{i-\tau}+\bar Z_i &\text{for } i\ge \nu+\tau.
\end{align*}
Finally, by multiplying the signal with the conjugate $e^{-j2\pi f i}$ and shifting time so that $Y_i = e^{-j2\pi f (i+\tau)}\bar Y_{i+\tau}$, we have post processed outputs given by
\begin{align*}
    Y_i &= Z_i &\text{for } i< \nu, \\
    Y_i &= hx_i+Z_i &\text{for } i\ge \nu,
\end{align*}
where $Z_i$ is the AWGN with $\Cc\Nc(0,1)$. In other words,
\begin{align*}
    Y_i &\sim \Cc\Nc(0,1) &\text{for } i< \nu, \\
    Y_i &\sim \Cc\Nc(hx_i,1) &\text{for } i\ge \nu.
\end{align*}
Then, with similar steps as in the scalar Gaussian example in Sec.~\ref{sec:Gaussian_example}, the tradeoff is given by
\begin{align*}
	(C,\Delta^\star) = \left(\log(1+P),\, |h|^2P\right).
\end{align*}

\subsection{MIMO Gaussian channels}
Consider the MIMO Gaussian sensing channels 
\begin{align*}
    \Yv_i &= G_0\xv_i+\Zv_i \quad \text{ for } i< \nu, \\
    \Yv_i &= G_1\xv_i+\Zv_i \quad \text{ for } i\ge \nu,
\end{align*}
and the communication channel $\tilde{\Yv}_i = \tilde{G}\xv_i+\tilde{\Zv}_{i}$, where $G_s$, $s=0,1$ and $\tilde{G}$ are the channel gain matrices for the sensing channels with state and the communication channel, respectively. Here, $\Zv_i, \tilde{\Zv}_{i}$ are independent vector Gaussian noise with distribution $\Nc(0, I)$, where $I$ is the identity matrix. Channel inputs are subject to a power constraint $\frac{1}{n} \sum_{i=1}^n \| \xv_i \|^2 \le P$.

We will first explore the pure QCD and pure communication cases. We begin by identifying $\Delta^\star$ for this case. When $\Xv = \xv$,
\begin{align*}
    D(p^{(1)} \|p^{(0)} |\Xv=\xv) &= \frac{1}{2} \xv^T \bar G^T \bar G \xv = \frac{1}{2} \xv^T\Gamma \xv.
\end{align*}
where $\Gamma := \bar G^T \bar G$ and $\bar G = G_1-G_0$. Letting $\Gamma = U \Lambda U^T$ be the singular value decomposition (SVD) of $\Gamma$ and $\bar{\xv}= U^T \xv$, 
\begin{align*}
    D(p^{(1)} \|p^{(0)} |\Xv=\xv) = \frac{1}{2} \xv^T\Gamma \xv = \frac{1}{2} \bar\xv^T\Lambda \bar\xv.
\end{align*}
Hence, the optimal QCD symbol for $\Delta^\star$ in \eqref{eq:delta_star} is $\bar{\xv} = [\sqrt{P},0,\ldots, 0]$ which leads to
\begin{align}
    \Delta^\star = \max_{\xv: \| \xv \|^2 \le P} D(p^{(1)} \|p^{(0)} |\Xv=\xv) = \frac{1}{2} \lambda_1 P,\label{eq:mimo_eq1}
\end{align}
where $\lambda_1$ is the largest singular value of $\Gamma$. In the following, we show that a Gaussian distribution can also achieve $\Delta^\star$, which will allow a positive communication rate. Note that since $\| U \xv \| = \| \xv \|$ for any unitary matrix $U$, 
\begin{align*}
    &D(p^{(1)} \|p^{(0)} | p_{\Xv} ) = \E ( D(p^{(1)} \|p^{(0)} | \Xv ) ) \\
    &= \frac{1}{2} \E \left( \Xv^T U \Lambda U^T \Xv \right) = \frac{1}{2} \E \left( \bar{\Xv}^T \Lambda \bar{\Xv} \right) \le \frac{1}{2} \lambda_1 P,
\end{align*}
where the equality is attained when $\bar{\Xv} \sim \Nc(0, \bar{\Sigma}_X)$ with $\bar{\Sigma}_X = \text{diag}([P, 0, \ldots, 0])$. It means that for QCD, an alternative optimal strategy is to send a vector Gaussian random code by allocating all the power to the sub-channel with maximum singular value. In a sense, the strategy translates to achieving maximum diversity gain for detection.

On the other hand, the communication rate for a given covariance matrix $\Sigma_X$ is given by
\begin{align*}
    I(\Xv; \tilde{\Yv}) &= \frac{1}{2}\log\left|I+\tilde{G} \Sigma_X \tilde{G}^T\right|.
\end{align*}
The optimal input distribution for the communication-only case is attained by beamforming with respect to SVD and waterfilling over the singular values of $\tilde{G}$~\cite{Telatar1999}.

This example entails several interesting observations in the tradeoff between communication and QCD. 
On the one hand, to achieve the best QCD detection delay, the transmitter should allocate all its power to the best subchannel with respect to the channel gain difference matrix $\bar G$. On the other hand, the best communication strategy is to multiplex over subchannels of $\tilde{G}$, i.e., distribute power across the subchannels of $\tilde{G}$ with a waterfilling policy. Compared to the single antenna case, the MIMO scenario has a more complicated tradeoff in that the mismatch in beamforming and power allocation due to the difference in $\bar G$ and $\tilde{G}$ results in a more delicate tradeoff involving diversity vs.~multiplexing gains.

\begin{figure}[!t]
	\vspace{-0em}
	\centering
 \resizebox{!}{13.5em}{
	\input{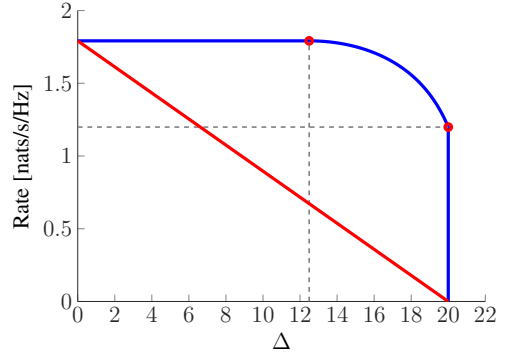}}
	\caption{Plot for the MIMO Gaussian example with $P=10$.}
	\label{fig:mimo_ex}
\end{figure}

We plot an example case for the MIMO setup in Fig.~\ref{fig:mimo_ex} where we assume power constraint $P=10$ and
\begin{align}
    \bar G = \begin{bmatrix}
        2 & 0 \\
        0 & 1 
    \end{bmatrix},\quad  \tilde{G} = \frac{1}{\sqrt{2}}\begin{bmatrix}
        1 & 1 \\
        1 & -1 
    \end{bmatrix}.
\end{align}

Note that $\bar G$ has a strictly maximal singular value while $\tilde{G}$ has two equal singular values. For optimal QCD, the optimal strategy is to allocate all power to the maximum singular value of $\bar G$ which achieves $\Delta^\star$, while the best strategy for communication is to allocate equal power to both antennas which achieves the capacity $C$. The plot in Fig.~\ref{fig:mimo_ex} shows $\mathscr{R}_\inn$ for the MIMO example. The red dots are the points $(C, \Delta(C))$ and $(R(\Delta^\star), \Delta^\star)$.

\section{Discussions}\label{sec:discussion}
This paper studies the information-theoretic tradeoff between communication rate and quickest change detection delay in the context of ISAC. Our goal in this work was to establish some firm foundations by taking essential initial steps towards determining the information-theoretic tradeoff. Nonetheless, several important open problems are still due, for instance, a tighter inner bound. The main difficulty in our strategy of applying Lem.~\ref{lem:Lai1998} to a generic converse is in the step that shows that the outputs produced from an {\em optimal} codebook satisfies~\eqref{eq:Lai_condition}. 
Since the QCD formulation requires good control over the empirical distribution of codeword induced outputs, developing a generic outer bound seems to involve analysis of the optimal code statistics with constraints, which is itself a widely active field of study, e.g.,~\cite{Shamai--Verdu1997, Polyanskiy--Verdu2014, Ding--Jaggi--Vatedka--Zhang2020}. A step in this direction would help us better understand the fundamental limits of joint communication and QCD.

The model considered in this work is useful in that it serves as a foundation for optimal simultaneous communication and QCD signal design principles. However, there are several important limitations to our bistatic model with a state-independent data channel. For instance, the case in which the state also impacts the communication channel $p(\yt|x,s)$ can be a model for the case when the physical phenomenon altering the state of the QCD channel also has impact on the communication channel. This type of channel model becomes more interesting when considering the monostatic model in Fig.~\ref{fig:monostatic_model}. In the monostatic model, the encoder can be interactive, which opens the possibility to encode with the knowledge of feedback, i.e., $x_i(m,y^{i-1})$. When coupled with the assumption $p(\yt|x,s)$, the adaptive signaling and state detection can also be used to aid communication. Whether an interactive encoder can improve the tradeoff performance is yet another interesting direction for future work.

\appendices

\section{Proof of Theorem~\ref{thm:theorem1}}\label{sec:proof_thm1}
In this section, we begin with the analysis on the achievable rate of constant subblock-composition codes (CSCCs) and then provide the analysis of the delay performance. Notice that it is sufficient to show that a pair $(I(p_X, p_{\Yt|X}), D(p^{(1)} \| p^{(0)} | p_X))$ is achievable for an arbitrary input distribution $p_X$. To this end, we fix an arbitrary $\epsilon > 0$ and show that $(I(p_X, p_{\Yt|X}) - 2\epsilon, D(p^{(1)} \| p^{(0)} | p_X)(1+\epsilon))$ is achievable. Let $L$ be a fixed integer that is large enough so that $p_X^{(L)} \in \Pc_L(\Xc)$ is ``close'' to $p_X$. Additional conditions will be stated throughout the proof. 

\subsection{Coding Rate Analysis} \label{subsec:rate_analysis}
Let CSCC$(L, p_X^{(L)})$ be the constant subblock-composition code of subblock length $L$ and composition type $p_X^{(L)} \in \Pc_L(\Xc)$. Note that the entire blocklength $n=kL$, where $k$ is the number of subblocks, and the blocklength will first be scaled with $k$ to obtain infinitely long codewords.

Let $r(L, p_X^{(L)})$ be a positive number defined by
    \begin{align*}
        r(L, p_X^{(L)}) &= \frac{(s(p_X^{(L)}) - 1)\log (2\pi L)}{2L}\\
        &+ \frac{1}{2L} \sum_{a: p_X^{(L)}(a) > 0} \log p_X^{(L)}(a) + \frac{u(L, p_X^{(L)})s(p_X^{(L)}) }{ 12L \ln 2},
    \end{align*}
where $s(p_X^{(L)})$ is the number of elements $x \in \mathcal{X}$ such that $p_X^{(L)}(x) > 0$ and $0 \le u(L, p_X^{(L)}) \le 1$ is some bounded function \cite[p.~25]{Csiszar--Korner2011}. Then, the coding rate of CSCC$(L, p_X^{(L)})$ is given as follows.
\begin{lemma}[Thm.~7 in \cite{Tandon--Motani--Varshney2016}]
    The code rate $R$ of the CSCC$(L, p_X^{(L)})$ with vanishing maximal block error probability as $n \to \infty$ is bounded by
    \begin{align*}
        I(p_X^{(L)}, p_{\Yt|X}) -r(L, p_X^{(L)}) \le R \le I(p_X^{(L)}, p_{\Yt|X}).
    \end{align*}
\end{lemma}
As $L$ is large and the mutual information is continuous in its input distribution, $p_X^{(L)}$ can be chosen close to $p_X$ so that $I(p_X, p_{\Yt|X}) -\epsilon < I(p_X^{(L)}, p_{\Yt|X})$. Also, the rate penalty term $r(L, p_X^{(L)})$ diminishes with $L$ since $r(L, p_X^{(L)}) = O(\frac{\log L}{L})$ and $p_X^{(L)}$ stays around $p_X$. Therefore, $r(L, p_X^{(L)}) < \epsilon$, which implies that 
$R > I(p_X, p_{\Yt|X}) - 2\epsilon$
is achievable.

\subsection{QCD Delay Analysis} \label{subsec:QCD_proof}
For the delay analysis, we require the following lemma. Notice that when $Z_1, Z_2, \ldots$ are all i.i.d., this lemma reduces to the original Wald's identity~\cite[Thm.~4.8.6]{Durrett2019}.
\begin{lemma}[Modified Wald's identity] \label{lem:mod_wald_identity}
	Let $Z_1, Z_2, \ldots$ be independent and $Z_2, Z_3, \ldots$ be identically distributed with finite mean. Then, for any stopping time $N$ that is finite with probability $1$, 
	\begin{align*}
		\E\left( \sum_{i=1}^N Z_i \right) = \E ( Z_2 ) \E(N) + (\E(Z_1) - \E(Z_2)) \P(1 \le N).
	\end{align*}
	Further, suppose that $|Z_i|<c$ for all $i$ with probability $1$. Then,
	\begin{align*}
		\E\left( \sum_{i=1}^N Z_i \right) \ge \E ( Z_2 ) \E(N) - 2c.
	\end{align*}
\end{lemma}
\begin{IEEEproof}[Proof of Lemma \ref{lem:mod_wald_identity}]
	Note that $\mathbf{1} \{ N \ge i \}  = 1 - \mathbf{1} \{ N \le i-1 \} $. Then, the right side only depends on $Z_1, Z_2, \ldots, Z_{i-1}$ by the definition of the stopping time, which also implies that the left side is independent of $Z_i$. Therefore,
    \begin{align*}
		&\E\left( \sum_{i=1}^N Z_i \right) = \E\left( \sum_{i=1}^{\infty} Z_i \mathbf{1}{ \{ i \le N \} } \right) = \sum_{i=1}^{\infty} \E ( Z_i \mathbf{1}{ \{ i \le N \} } ) \\
        &\stackrel{(a)}{=} \sum_{i=1}^{\infty} \E ( Z_i ) \E ( \mathbf{1}{ \{ i \le N \} } ) = \sum_{i=1}^{\infty} \E ( Z_i ) \P ( i \le N ) \\
		&= \left( \sum_{i=1}^{\infty} \E ( Z_2 ) \P ( i \le N ) \right) + (\E(Z_1) - \E(Z_2)) \P( 1 \le N ) \\
		&\stackrel{(b)}{=} \E ( Z_2 ) \E(N) + (\E(Z_1) - \E(Z_2)) \P( 1 \le N ),
	\end{align*}
	where $(a)$ follows from the independence, and $(b)$ follows from $\E(N) = \sum_{i=1}^{\infty} \P(i \le N)$. The first claim is proved.
	
The second claim is proved as follows:
\begin{align*}
    \E\left( \sum_{i=1}^N Z_i \right) &= \E ( Z_2 ) \E(N) + (\E(Z_1) - \E(Z_2)) \P( 1 \le N ) \\
    &\ge \E ( Z_2 ) \E(N) - \E(|Z_1|)  - \E(|Z_2|) \\
    &\ge \E ( Z_2 ) \E(N) - 2c. & \IEEEQEDhere
\end{align*}
\end{IEEEproof}

We are now ready to analyze the delay asymptotics. Consider a sequence of CSCCs $\Ccn$, $n=kL$, with $k \to \infty$ and $L$ being fixed. 
For QCD, we use the subblock-CuSum (SCS) test defined in~\eqref{eq:SCS} and~\eqref{eq:NSCS}. By its construction, $W_i$ only changes at $i=jL$, and therefore, $N_\SCS \in \{jL\}_{j \in \mathbb{N}}$ or $N_\SCS = n+1$ if the threshold has not been reached. Noting that the probability of $N_\SCS = n+1$ can be made arbitrarily small by taking a large enough $n$, we ignore such event in the following analysis.

A key idea that connects our CSCC-based QCD and the standard QCD for i.i.d.~observations is from the fact that by properly permuting transmission symbols of any CSCC codeword, each subblock of codewords can be thought of as a repetition of $\xv_1(1)$, the first subblock of $x^n(1)$. Hence, for any $x^n(m)$, the permuted observation vectors are i.i.d.~observations from $\xv_1(1)$. To be precise, for the $j$-th subblock of the $m$-th codeword, let $\sigma_{j,m}$ be the permutation such that $\sigma_{j,m}(\xv_j(m)) = \xv_1(1)$, which always exists since all subblocks in codewords are of the same composition. For observations, let $\yv_{j}'(m):= \sigma_{j,m}(\yv_j(m))$ be a vector obtained from observations in the $j$-th subblock under permutation $\sigma_{j,m}$. Then, the distributions of subblocks of observations can be categorized into three classes: If the $j$-th subblock is in the prechange period, the $j$-th random vector after permutation, $\Yv_j'(m)$, is identically distributed to the unpermuted observation vector $\Yv_1(1)$ obtained from $\xv_1(1)$, i.e.,
\begin{align*}
    p^{(0)}_{\Yv_1|\xv_1} (\yv_j') = \prod_{i=1}^L p^{(0)}_{Y|X}(y_{(j-1)L+i}'|x_i(1)).
\end{align*}
Similarly, if the subblock is strictly after the change point, then the permuted observation vectors are distributed by
\begin{align*}
    p^{(1)}_{\Yv_1|\xv_1} (\yv_j') = \prod_{i=1}^L p_{Y|X}^{(1)}(y_{(j-1)L+i}'|x_i(1)).
\end{align*}
However, if the change is in the middle of the subblock, the permuted observations' distribution is not $p^{(1)}_{\Yv_1|\xv_1}$ nor $p^{(0)}_{\Yv_1|\xv_1}$.

Let $N_{\SCS}':= N_{\SCS}/L = \inf \{ j \ge 1: W_{jL} \ge b \}$ be the equivalent stopping time, which sees the following QCD problem after proper permutations:
\begin{align*}
    \Yv_j' &\sim p^{(0)}_{\Yv_1|\xv_1} ~~~ \text{if } jL < \nu \\
    \Yv_j' &\sim \tilde{p}_{\Yv_1|\xv_1} ~~~ \text{if } (j-1)L + 1 < \nu \le jL \\
    \Yv_j' &\sim p^{(1)}_{\Yv_1|\xv_1} ~~~ \text{if } \nu \le (j-1)L + 1 
\end{align*}
where $\xv_1 = \xv_1(1)$ is the reference transmission vector and $\tilde{p}_{\Yv_1|\xv_1}$ is some distribution that depends on the location of $\nu$ and $\xv_1$. The first and last intervals are for the vectors that are completely within the pre-change and post-change periods, respectively. The second interval is for the vector that contains the change point $\nu$. Note that vectors $\Yv_1', \Yv_2', \ldots$ are independent conditioned on a transmitted codeword since the channel is memoryless.

The QCD setup described above differs from standard cases if the change occurs in between a subblock, i.e., $\nu$ is in the middle of a subblock. The precise distribution of $\tilde{p}_{\Yv_1|\xv_1}$ will be irrelevant in the proof steps, however, what is relevant is the fact the it is not necessarily distributed by the alternative hypothesis $p^{(1)}_{\Yv_1|\xv_1}$. Fortunately, even in this scenario, one can extend Lorden's analysis to obtain bounds on the performance in terms of sequential probability ratio tests (SPRTs)~\cite[Thm.~2]{Lorden1971}.

To be precise, consider the following one-sided SPRT with stopping time $N_{\SPRT}'$. Let $\Yv_1', \Yv_2', \ldots$ be independent observations conditioned on the transmitted codeword such that
\begin{align*}
    \text{if } &H_0: ~ \Yv_j' \sim p^{(0)}_{\Yv_1|\xv_1} ~~ \text{for all } j, \\
    \text{if } &H_1: ~ \Yv_1' \sim \tilde p_{\Yv_1|\xv_1}, ~~ \Yv_j' \sim p^{(1)}_{\Yv_1|\xv_1} ~~ \text{for } j \ge 2, \\
    N_{\SPRT}' &:= \inf \left\{ j \ge 1: \log \frac{\tilde p_{\Yv_1|\xv_1}}{p^{(0)}_{\Yv_1|\xv_1} }(\Yv_1')\right.\\
    &\qquad\qquad\qquad\quad \left.+ \sum_{k=2}^j \log \frac{ p^{(1)}_{\Yv_1|\xv_1}  }{ p^{(0)}_{\Yv_1|\xv_1} } (\Yv_k') \ge b \right\},
\end{align*}
where $H_0$ and $H_1$ are the null and alternative hypotheses, respectively.
As above, it is slightly different from the standard SPRT in that the first observation under the alternative hypothesis does not follow $p^{(1)}_{\Yv_1|\xv_1}$. Further, note that $p^{(0)}_{\Yv_1|\xv_1}, p^{(1)}_{\Yv_1|\xv_1}$ remain unchanged regardless of the choice of codewords if proper permutations are available. 

Note that our $N_{\SCS}'$ can be interpreted as a set of SPRTs where $N_{\SPRT, j}'$ denotes the SPRT applied to subblocks $\Yv_j', \Yv_{j+1}', \ldots$, by
\begin{align}
    N_{\SCS}' = \min \{ N_{\SPRT, j}' + j - 1, j = 1,2,\ldots \}. \label{eq:CS_SPRT_relation}
\end{align}
Let $\nu'\in[1:k]$ be the subblock index that contains $\nu$. Then, the analysis for the standard i.i.d.~QCD/SPRT problem where $\tilde p = p^{(1)}$ \cite[Thm.~2]{Lorden1971} can be extended to our vector-symbol QCD/SPRT problem that has $\tilde p \neq p^{(1)}$, by which we have the following lemma.
\begin{lemma}\label{lem:lorden_SPRT}
    Consider a sequence of CSCC codebooks $\Ccn$. If 
    \begin{align*}
        \limsup_{n \to \infty} \max_{x^n \in \Ccn} \P( N_{\SPRT}' < \infty | x^n, H_0) \le \alpha',
    \end{align*}
    then for our QCD problem,
    \begin{align}
        &\limsup_{n \to \infty} \max_{x^n \in \Ccn} \E_{\infty} (N_{\SCS}' | x^n) \ge \frac{1}{\alpha'} \label{eq:lorden_1} 
    \end{align}
    and
    \begin{align}
        &\sup_{\nu' \ge 1} \limsup_{n \to \infty} \max_{x^n \in \mathcal{C}_n} \esssupe_{Y^{\nu'-1}} \E_{\nu}((N_{\SCS}'-\nu'+1)^+|x^n, \Yv^{\nu'-1})\nn\\
        &\le \max\{ \limsup_{n \to \infty} \max_{x^n \in \Ccn} \E_1 (N_{\SPRT,1}' | x^n), \E_2(N_{\SPRT, 2}') \}. \label{eq:lorden_2}
    \end{align}
\end{lemma}
\begin{IEEEproof}[Proof of Lemma \ref{lem:lorden_SPRT}]
    The proof of \eqref{eq:lorden_1} follows exactly the same steps as in the proof of\cite[Thm.~2]{Lorden1971} since the vectors of observations with proper permutation are i.i.d.~under $H=0$. For the proof of \eqref{eq:lorden_2}, by \eqref{eq:CS_SPRT_relation}, if $\nu' = 1$, for every $x^n$,
    \begin{align*}
        \E_1 (N_\SCS' | x^n) &\le \E_1 (N_{\SPRT,1}' | x^n),
    \end{align*}
    and if $\nu' \ge 2$,
    \begin{align*}
         \E_{\nu'} ( (N_\SCS'-\nu'+1)^+ | x^n, \Yv^{\nu'-1}) &\le \E_{\nu'} (N_{\SPRT, \nu'}' | x^n, \Yv^{\nu'-1})\\
        &= \E_{\nu'} (N_{\SPRT,\nu'}') \\
        &= \E_{2} (N_{\SPRT,2}'),
    \end{align*}
    where the first equality holds since permuted vectors are i.i.d.~if $\nu' \ge 2$ and are independent of the past ones. Hence, the claim is proved.
\end{IEEEproof}

Now, the problem reduces to bounding $\E_1 (N_{\SPRT,1}' | x^n)$ and $\E_2(N_{\SPRT, 2}')$. Following Wald's SPRT analysis~\cite{Wald1947}, the false alarm probability is $\P(\text{false alarm}) := \limsup_{n \to \infty} \max_{x^n \in \Ccn} \P( N_{\SPRT}' < \infty | x^n, H_0 ) \le e^{-b}$, where $b$ is the threshold in $N_\SPRT'$. Also, note that since $|\mathcal{X}|, |\mathcal{Y}|$ are finite, 
\begin{align}
    \max_{x, y} \left| \log \frac{ p^{(1)}(y|x) }{ p^{(0)}(y|x) }  \right| =: \gamma < \infty. \label{eq:finite_LR}
\end{align}
Let 
\begin{align*}
S_j := \log \frac{\tilde p_{\Yv_1|\xv_1}}{p^{(0)}_{\Yv_1|\xv_1} }(\Yv_1') + \sum_{k=2}^j \log \frac{ p^{(1)}_{\Yv_1|\xv_1}  }{ p^{(0)}_{\Yv_1|\xv_1} } (\Yv_k').
\end{align*}
Then, $\E_1 (N_{\SPRT, 1}' | x^n)$ can be bounded by the modified Wald's identity in Lem.~\ref{lem:mod_wald_identity} as follows: for a large enough $n$ and $x^n\in\Ccn$,
\begin{align}
    &b(1+o(1)) \nn \\
    &\stackrel{(a)}{=} \E_1 (S_{N_{\SPRT,1}'} | x^n ) \nn\\
    &\stackrel{(b)}{=} \E_1(N_{\SPRT,1}' | x^n ) \E_{1}\left( \log \frac{ p^{(1)}_{\Yv_1|\xv_1} }{ p^{(0)}_{\Yv_1|\xv_1} } (\Yv_2') \right) \nn\\
    &\quad \quad \quad+  \E_{1} \left( \log \frac{ \tilde p_{\Yv_1|\xv_1} }{ p^{(0)}_{\Yv_1|\xv_1} } (\Yv_1') \Bigg| x^n \right)\P(1 \le N_{\SPRT,1}' | x^n) \nn\\
    &\quad \quad \quad - \E_{1}\left( \log \frac{ p^{(1)}_{\Yv_1|\xv_1} }{ p^{(0)}_{\Yv_1|\xv_1} } (\Yv_2') \right)  \P(1 \le N_{\SPRT,1}' | x^n) \nn\\
    &\stackrel{(c)}{\ge} \E_1(N_{\SPRT,1}' | x^n ) \E_{1}\left( \log \frac{ p^{(1)}_{\Yv_1|\xv_1} }{ p^{(0)}_{\Yv_1|\xv_1} } (\Yv_2') \right) - 2\gamma L \nn\\
    &= \E_1( N_{\SPRT,1}' | x^n ) D\left( p^{(1)}_{\Yv_1\xv_1} \big\| p^{(0)}_{\Yv_1|\xv_1} \right) -2\gamma L, \label{eq:sprt_bnd1}
\end{align}
where $(a)$ follows since the overshoot is negligible if $b$ is large, and $(b)$, $(c)$ follow from the modified Wald's identity with \eqref{eq:finite_LR}. Then, \eqref{eq:sprt_bnd1} implies that for large enough $n$,
\begin{align}
    \E_1(N_{\SPRT,1}' | x^n) \le \frac{ b(1+o(1)) + 2\gamma L }{ D\left( p^{(1)}_{\Yv_1|\xv_1} \big\| p^{(0)}_{\Yv_1|\xv_1} \right)  },
\end{align}
which is independent of what $x^n$ was transmitted.
Also, Wald's original identity~\cite[Thm.~4.8.6]{Durrett2019} computes $\E_2(N_{\SPRT, 2}')$ as follows:
\begin{align*}
    \E_2(N_{\SPRT,2}') \le \frac{ b(1+o(1)) }{ D\left( p^{(1)}_{\Yv_1|\xv_1}\big\| p^{(0)}_{\Yv_1|\xv_1} \right) }.
\end{align*}

Letting $\alpha' = e^{-b}$, we have $b = |\log \alpha'|$; in other words, $\E_{\infty} (N_{\SCS}' | x^n) \ge \frac{1}{\alpha'}$ for large enough $n$ and 
\begin{align*}
    &\sup_{\nu' \ge 1} \limsup_{n \to \infty} \max_{x^n \in \mathcal{C}_n} \esssup_{Y^{\nu'-1}} \E_{\nu'}((N_{\SCS}'-\nu'+1)^+|x^n, \Yv^{\nu'-1})\\
    &\quad \le \frac{ |\log \alpha'| (1+o(1)) + 2\gamma L }{ D\left( p^{(1)}_{\Yv_1|\xv_1}\big\| p^{(0)}_{\Yv_1|\xv_1} \right) }.
\end{align*}

Finally, using two facts that $N_{\SCS} = L \cdot N_{\SCS}'$ and $D\left( p^{(1)}_{\Yv_1|\xv_1}\big\| p^{(0)}_{\Yv_1|\xv_1} \right) = L \cdot D(p^{(1)} \| p^{(0)} | p_X^{(L)})$ by the property of the KL divergence for product distributions, one can convert it back to the original QCD bounds given by
$\E_\infty (N_{\SCS} | x^n) \ge \frac{L}{\alpha'}$
and
\begin{align*}
    &\sup_{\nu \ge 1} \limsup_{n \to \infty} \max_{x^n \in \mathcal{C}_n} \esssup_{Y^{\nu-1}} \E_{\nu}((N_{\SCS}-\nu+1)^+|x^n, Y^{\nu-1})\\
    &\quad \le \frac{ L\cdot|\log \alpha'| (1+o(1)) + 2\gamma L^2 }{ L \cdot D(p^{(1)} \| p^{(0)} | p_X^{(L)}) } \\
    &\quad = \frac{ |\log \alpha'| (1+o(1)) + 2\gamma L }{ D(p^{(1)} \| p^{(0)} | p_X^{(L)}) }.
\end{align*}
By taking $\alpha := \frac{\alpha'}{L} \Leftrightarrow \log \alpha' = \log \alpha + \log L$,
\begin{align*}
    \E_\infty (N_{\SCS} | x^n) \ge \frac{1}{\alpha} 
\end{align*}
and
\begin{align*}
    \WADD(N_{\SCS}) \le \frac{ |\log \alpha| (1+o(1)) + \log L + 2\gamma L }{ D(p^{(1)} \| p^{(0)} | p_X^{(L)}) }. 
\end{align*}
As $L$ is fixed but large, one can express the KL divergence as $D(p^{(1)} \| p^{(0)} | p_X^{(L)}) < D(p^{(1)} \| p^{(0)} | p_X) / (1+\epsilon)$. Then, for small enough $\alpha$, we have 
$\E_\infty (N_{\SCS} | x^n) \ge \frac{1}{\alpha}$
and
\begin{align*}
    \WADD(N_{\SCS}) \le \frac{ |\log \alpha| (1+o(1))(1+\epsilon) }{ D(p^{(1)} \| p^{(0)} | p_X) } =: \Ds.
\end{align*}
It proves the claim that
\begin{align*}
    \Delta = \lim_{\alpha \to 0} \frac{ |\log \alpha| }{\Ds} = D(p^{(1)} \| p^{(0)} | p_X) (1+\epsilon).
\end{align*}	
As $\epsilon$ is arbitrary, we proved the claim.

\section{Proof of Theorem~\ref{thm:theorem2}}\label{sec:proof_thm2}
Before discussing the proof of Thm.~\ref{thm:theorem2}, we state the following result by Lai~\cite{Lai1998}, where observations are generic sequences that are not necessarily independent nor identically distributed. Here, $P^{(\nu)}$ denotes the probability measure when the change occurs at $\nu$.
\begin{lemma}[{\cite[Thm.~1]{Lai1998}}]\label{lem:Lai1998}
	Suppose that a QCD receiver has observations $Y_1, Y_2, \ldots$ that satisfy
	\begin{align}
		\tilde P^{(\nu)} \left( \max_{t \le T} \frac{1}{T} \sum_{i=\nu}^{\nu + t-1} \log \frac{p^{(1)}(Y_i)}{p^{(0)}(Y_i)} \ge (1+\delta) c ~ \bigg| ~ Y^{\nu - 1} \right) = 0  \label{eq:Lai_condition}
	\end{align}
for any $\delta > 0 $ with some positive constant $c > 0$, where $\tilde  P^{(\nu)}(\cdot) = \lim_{T \to \infty} \sup_{\nu \ge 1} \esssupe_{Y^{\nu - 1}} P^{(\nu)}(\cdot)$. Then, for any stopping rule $N$ such that $\frac{1}{\E_{\infty}(N)} \le \alpha$,
	\begin{align*}
		&\sup_{\nu \ge 1} \esssupe_{Y^{\nu - 1}} \E_{\nu}((N-\nu+1)^+|Y^{\nu-1}) \\
  &\quad \ge |\log \alpha|\left( {\frac{1}{c} + o(1)} \right) ~~~ \text{as $\alpha \to 0$.}
	\end{align*}
\end{lemma}

Since Lai's result states a lower bound on the QCD delay for more general observations beyond the i.i.d.~case, it is sufficient to show that a sliding-window typical sequence of codes yields QCD observations that satisfy the condition \eqref{eq:Lai_condition}. Let $c = D(p^{(1)} \| p^{(0)} | p_X)$ and $L_0$ be the interval length satisfying~\eqref{eq:sliding_typical} for some $\epsilon > 0$. To give some preparation for what follows, we note that $L_0$ serves as an interval length such that $x^{L_0}$ is typical, i.e., it plays the same role as $L$ in CSCCs. Also, $K_0 L_0$ will serve as an interval length such that $Y^{K_0 L_0}$ is typical. To be clear, intervals of length $K_0 L_0$ and $L_0$ are denoted by subintervals and microintervals, respectively.

Suppose that $T$ in~\eqref{eq:Lai_condition} grows as a multiple of $K_0 L_0$, i.e., $T = k(K_0 L_0), k \in \mathbb{N}$, where $K_0$ is an integer that will be specified later. Then, for a sequence of $\Ccn$, the condition to show for our setting is that for any $\delta > 0$,
\begin{align}
    \bar P^{(\nu)} \left( \max_{t \le k K_0 L_0} \frac{1}{k K_0 L_0} \sum_{i=\nu}^{\nu + t-1} Z_i \ge (1+\delta) c ~ \bigg| ~ \xv_{\nu}\right) = 0, \label{eq:claim_to_show}
\end{align}
where $Z_i := \log \frac{p^{(1)}(Y_i|x_i)}{p^{(0)}(Y_i|x_i)}$, $\xv_\nu=x_{\nu}^{\nu+k K_0 L_0-1}$,
\begin{align*}
    \bar P^{(\nu)}(\cdot)=\lim_{k \to \infty} \sup_{\nu \ge 1} \limsup_{n \to \infty} \max_{x^n \in \mathcal{C}_n} P^{(\nu)}(\cdot).
\end{align*} 
Here, $Y^{\nu-1}$ is dropped since observations from a memoryless channel are independent of the past observations conditioned on $\xv_\nu$, and thus, $Z_i, i \ge \nu$ is independent of the past.

Note that $[\nu: \nu+t-1]$ consists of $ \lfloor t/K_0 L_0 \rfloor $ subintervals of length $K_0 L_0$ and residuals of length less than $K_0 L_0$, the effect of which becomes negligible as $ \lfloor t/K_0 L_0 \rfloor $ grows, i.e.,
\begin{small}
\begin{align*}
    &\frac{1}{k K_0 L_0} \sum_{i=\nu}^{\nu + t-1} Z_i \\
    &= \frac{1}{k} \cdot \frac{1}{K_0 L_0} \left( \sum_{j=1}^{ \lfloor t/K_0 L_0 \rfloor } \sum_{i=\nu + (j-1)K_0 L_0}^{\nu + j K_0 L_0 - 1} Z_i + \sum_{i=\nu + \lfloor t/K_0 L_0 \rfloor L_0}^{\nu + t - 1} Z_i \right) \\
    &= \frac{1}{k} \cdot \frac{1}{K_0 L_0} \sum_{j=1}^{ \lfloor t/K_0 L_0 \rfloor } \sum_{i=\nu + (j-1)K_0 L_0}^{\nu + j K_0 L_0 - 1} Z_i + o(1) ~~~ \text{as } k \to \infty.
\end{align*}
\end{small}

Fix an arbitrary $\delta > 0$ and consider the first subinterval of length $K_0 L_0$, i.e., $i \in [\nu: \nu + K_0 L_0 - 1]$, that consists of $K_0$ micro-intervals. Then, the sum over this subinterval can be rewritten by dividing it into $|\mathcal{X}|$ partial sums according to $x$ and then further dividing it into $|\mathcal{Y}|$ sums,
\begin{align*}
    &\frac{1}{K_0 L_0} \sum_{i=\nu}^{\nu + K_0 L_0 - 1} \log \frac{ p^{(1)}(y_i|x_i) }{p^{(0)}(y_i|x_i)}\\
    &= \sum_{x \in \mathcal{X}} \frac{ n_x }{K_0 L_0} \left( \frac{1}{ n_x } \sum_{i: x_i = x} \log \frac{ p^{(1)}(y_i|x) }{p^{(0)}(y_i|x)} \right) \\
    &= \sum_{x \in \mathcal{X} } \frac{n_x}{K_0 L_0} \left( \sum_{y \in \mathcal{Y}} \frac{n_{y|x}}{n_x} \log \frac{ p^{(1)}(y|x) }{ p^{(0)}(y|x)} \right),
\end{align*}
where $n_x := | \{i \in [\nu: \nu+K_0 L_0 - 1]: x_i = x\} |$ and $n_{y|x} := |\{i \in [\nu: \nu+K_0 L_0 - 1]: x_i = x, y_i = y \}| $. Since $x^n$ is sliding-window typical, $|\frac{n_a}{K_0 L_0} - p(a)| \le \epsilon p(a)$ for all $a \in \mathcal{X}$ by the definition of $L_0$. Furthermore, by the conditional typicality lemma~\cite{El-Gamal--Kim2011}, if $K_0$ is sufficiently large, $y_{\nu}^{\nu+K_0 L_0-1}$ is conditionally $\epsilon'$-typical with high probability. In other words, for any $x \in \mathcal{X}, y \in \mathcal{Y}$, it holds that $|\frac{n_{y|x}}{n_x} - p^{(1)}(y|x)| \le \epsilon'p^{(1)}(y|x)$ with probability greater than $1- \delta'(\epsilon')$.

Combining the above for jointly typical sequences, i.e., with probability greater than $1- \delta'(\epsilon')$,
\begin{align*}
    &\frac{1}{K_0 L_0} \sum_{i=\nu}^{\nu + K_0 L_0 - 1} \log \frac{ p^{(1)}(y_i|x_i) }{p^{(0)}(y_i|x_i)}\\
    &\le \sum_{x \in \mathcal{X} } (1+\epsilon) p_X(x) \left( \sum_{y \in \mathcal{Y}} (1+\epsilon') p^{(1)}(y|x) \log \frac{ p^{(1)}(y|x) }{ p^{(0)}(y|x)} \right) \\
    &= (1+\epsilon)(1+\epsilon') D( p^{(1)} \| p^{(0)} | p_X ).
\end{align*}
On the other hand, for atypical observation sequences, i.e., with probability less than $\delta'(\epsilon')$,
\begin{align*}
    \frac{1}{K_0 L_0} \sum_{i=\nu}^{\nu + K_0 L_0 - 1} \log \frac{ p^{(1)}(y_i|x_i) }{p^{(0)}(y_i|x_i)} &\le \gamma,
\end{align*}	
where $\gamma$ is defined in \eqref{eq:finite_LR}. Then, one can verify that by taking $K_0$ appropriately so that
\begin{align}
    \mu_1 &:= \E \left( \frac{1}{K_0 L_0} \sum_{i=\nu}^{\nu + K_0 L_0 - 1} \log \frac{ p^{(1)}(Y_i|x_i) }{p^{(0)}(Y_i|x_i)} \right) \nn \\
    &\le (1+\epsilon'') D( p^{(1)} \| p^{(0)} | p_X ) \nn \\
    &< (1+\delta) D( p^{(1)} \| p^{(0)} | p_X ). \label{eq:K0_condition}
\end{align}

Next, noting that other subintervals are independent conditioned on $x^n$, we apply the same argument to other subintervals. To simplify notation, define $A_j$, $j \in [1:k]$, such that
\begin{align*}
    A_j := \frac{1}{K_0 L_0} \sum_{i=\nu + (j-1)K_0 L_0}^{\nu + j K_0 L_0 - 1} \log \frac{ p^{(1)}(Y_i|x_i) }{p^{(0)}(Y_i|x_i)} - \mu_j,
\end{align*}
where $\mu_j$ are positive values defined in a similar way to $\mu_1$. Note that $A_j$ has zero mean and finite variance. Then, the probability in \eqref{eq:claim_to_show} can be rewritten as
\begin{align*}
    P^{(\nu)} \left( \max_{t' \le k} \sum_{j=1}^{t'} (A_j + \mu_j) \ge k (1+\delta) D + o(k) ~ \bigg| ~ \xv_\nu \right),
\end{align*}
where $D=D( p^{(1)} \| p^{(0)} | p_X )$. Then,
\begin{align*}
    &P^{(\nu)} \left( \max_{t' \le k} \sum_{j=1}^{t'} (A_i + \mu_j) \ge k (1+\delta) D + o(k) ~ \bigg| ~ \xv_\nu \right) \\
    &\le P^{(\nu)} \left( \max_{t' \le k} \sum_{j=1}^{t'} A_i \ge k (1+\delta) D - \sum_{j=1}^{t'} \mu_j + o(k) ~ \bigg| ~ \xv_\nu \right) \\
    &\stackrel{(a)}{\le} P^{(\nu)} \left( \max_{t' \le k} \sum_{j=1}^{t'} A_i \ge k (\delta - \epsilon'') D + o(k) ~ \bigg| ~ \xv_\nu \right) \\
    &\le P^{(\nu)} \left( \max_{t' \le k} \left| \sum_{j=1}^{t'} A_i \right| \ge k (\delta - \epsilon'') D + o(k) ~ \bigg| ~ \xv_\nu \right) \\
    &\stackrel{(b)}{\le} \frac{ k \max_j \Var(A_j) }{ k^2 (\delta - \epsilon'')^2 D^2( p^{(1)} \| p^{(0)} | p_X ) + o(k^2) } \\
    &= O(1/k) \to 0,
\end{align*}
where $(a)$ follows from \eqref{eq:K0_condition} and $(b)$ follows from Kolmogorov's maximal inequality~\cite[Thm.~2.5.5]{Durrett2019}. Since the bounds do not depend on $\nu$ nor a specific choice of $x^n \in \Ccn$, we have proved \eqref{eq:claim_to_show} for $T = k K_0 L_0$. The proof for $T \ne k K_0 L_0$ is immediate since the effect of residuals is negligible.	

\section{Proof of Theorem~\ref{thm:theorem3}}\label{sec:proof_thm3}
Suppose that $(R, \Delta)$ such that $R > 0$ and $\Delta > 0$ is achievable by a sequence of sliding-window typical codes. Then, by the relation of $\Delta$ and $\WADD$, for any $\epsilon > 0$, there exists a stopping rule $N^\star$ such that for sufficiently small $\alpha$ and sufficiently large $n$,
\begin{align}
	\Delta \le \frac{|\log \alpha|}{\WADD(N^\star)}(1 + \epsilon)  \Rightarrow  \WADD(N^\star) \le \frac{|\log \alpha|}{ \Delta } (1 + \epsilon). \label{eq:bound3_1}
\end{align}
Also, from Thm.~\ref{thm:theorem2}, if the sliding-window typical codes is with $p_X$, then for any stopping rule $N$,
\begin{align}
	\WADD(N) \ge |\log \alpha| \left( \frac{1}{ D( p^{(1)} \| p^{(0)} | p_X ) } + o(1) \right).\label{eq:bound3_2}
\end{align}
Combining \eqref{eq:bound3_1} and~\eqref{eq:bound3_2}, we can conclude that $p_X$ sliding-window typical codes can achieve $\Delta$ only if
\begin{align}
	\Delta \le D(p^{(1)} \| p^{(0)} | p_X). \label{eq:px_condition2}
\end{align}

Viewing the KL divergence for each input symbol as an input cost, i.e., $c(x) := D(p^{(1)} \| p^{(0)} | x), x \in \mathcal{X}$, \eqref{eq:px_condition2} can be treated as input cost constraint on the code:
$D(p^{(1)} \| p^{(0)} | p_X) = \E_{p_X}[c(X)] \ge \Delta$.
Hence, finding an upper bound on the coding rate with the cost constraint gives an upper bound on $R_{\pi}(\Delta)$, where
\begin{align*}
	R_{ \pi }(\Delta) &:= \sup \{R: (R, \Delta') \text{ is achievable using a }\\
 &\qquad \text{sliding-window typical code for some } \Delta' \le \Delta\}.
\end{align*}
To this end, define
$C(\Delta) := \max_{p_X: \E(c(X)) \ge \Delta} I(X;\Yt).$
Note that $C(\Delta)$ is nonincreasing, concave, and continuous in $\Delta$. Since our sequence of codes is sliding-window typical, for any $\delta > 0$, there exists a large enough $n_0(\delta)$ such that
\begin{align*}
	\frac{1}{n_0} \sum_{i=1}^{n_0} D(p^{(1)} \| p^{(0)} | x_i) &\ge D(p^{(1)} \| p^{(0)} | p_X) - \delta\\
    &= \E_{p_X}[c(X)]- \delta \\
    &\ge \Delta - \delta ~~~ \text{for all $x^{n_0} \in \Cc^{(n_0)}$}.
\end{align*}

By standard converse proof steps using Fano's inequality and the data processing inequality, if $n \ge n_0$,
\begin{align*}
	nR_{\pi} &\le I(X^n;\Yt^n) + n \epsilon_n \\
    &\le \sum_{i=1}^n I(X_i;\Yt_i) + n \epsilon_n \\
	&\stackrel{(a)}{\le} \sum_{i=1}^n C( \E(c(X_i)) ) + n \epsilon_n \\
    &\stackrel{(b)}{\le} n C \left( \frac{1}{n} \sum_{i=1}^n \E(c(X_i)) \right) + n \epsilon_n \\
	&\stackrel{(c)}{\le} n C(\Delta - \delta) + n \epsilon_n \\
    &\stackrel{(d)}{\le} n C(\Delta) + n \epsilon_n + \epsilon'(\delta) \\
	&= n \max_{p_X: D(p^{(1)} \| p^{(0)} | p_X) \ge \Delta} I(X;\Yt) + n \epsilon_n + \epsilon'(\delta),
\end{align*}
where step $(a)$ follows from the definition of $C(\cdot)$ and steps $(b)$--$(d)$ follow since $C(\cdot)$ is concave, nonincreasing, and continuous. Here, $\epsilon'$ is a quantity that depends on $\delta$ and vanishes if $\delta \to 0$. We have derived the upper bound that coincides with $\mathscr{R}_\inn$, which completes the proof.

\section{Blahut--Arimoto Algorithm}\label{app:BA}
We present the Blahut--Arimoto algorithm for the achievable rate region in Thm.~\ref{thm:theorem1}, i.e., numerical evaluation of $\mathscr{R}_\inn$ in~\eqref{eq:BA_objective}. The key idea of the original Blahut--Arimoto algorithm \cite{Blahut1972, Arimoto1972} is that the maximum mutual information can be found by an iterative optimization between input and posterior distributions, $r(x), q(x|\yt)$, respectively. 

For our setting, the objective function to maximize is written as
\begin{align*}
    J(r, q) &= \sum_{x}\sum_{\yt} r(x)p(\yt|x)\log\frac{q(x|\yt)}{r(x)}+\lambda \sum_{x}r(x)c(x),
\end{align*}
where $\lambda\in[0,\infty)$ and $c(x) = D(p^{(1)} \| p^{(0)} |X = x)$.
Then, the optimal input distribution $r^\star(x)$ for each $\lambda$ can be found by iterating the following two equations alternately with $r(x)$ being randomly initialized.
\begin{enumerate}
    \item For a fixed $r(x)$, $J(r, q)$ is maximized by
\begin{align*}
    q(x|\yt) = \frac{r(x)p(\yt|x)}{\sum_{x'} r(x')p(\yt|x')}.
\end{align*}

    \item For a fixed $q(\yt|x)$, $J(r, q)$ is maximized by
\begin{align*}
    r(x) &= \frac{\exp\left(\sum_{\yt}p(\yt|x) \log q(x|\yt) + \lambda c(x)\right)}{\sum_{x'} \exp\left(\sum_{\yt}p(\yt|x') \log q(x'|\yt) + \lambda c(x')\right)}.
\end{align*}
\end{enumerate}

\end{document}